\documentclass[apj]{emulateapj}
\usepackage{apjfonts}
\usepackage{lscape,graphicx,longtable}
\usepackage{amsbsy}
\newcommand{\cmjj}{\mbox{${\rm cm^{-2}}$}}
\newcommand{\hI}{\mbox{${\rm H\,I}$}}

\newcommand{\lya}{\mbox{${\rm Ly}\alpha$}}

\newcommand{\apg}{\gtrsim}
\newcommand{\apll}{\lesssim}
\newcommand{\etal}{\ensuremath{\mbox{et~al.}}}
\newcommand{\hmsol}{\mbox{$h^{-1}\,{\rm M}_\odot$}}

\providecommand{\kms}{\,\ensuremath{\rm{km\,s}^{-1}}}

\newcommand{\ewr}{\mbox{$W_r(2796)$}}

\shorttitle{Extended Cool Gas around Galaxies}
\shortauthors{Chen \etal}

\begin{document}

\slugcomment{To appear in the Astrophysical Journal 2010 May 10 v714 issue}
%\slugcomment{Draft Version on 23 December 2009}

\title{An Empirical Characterization of Extended Cool Gas Around Galaxies Using Mg\,II Absorption Features$^1$}
%\title{An Empirical Characterization of Extended Mg\,II Absorbing Gas around Galaxies$^1$}

\author{Hsiao-Wen Chen\altaffilmark{2}, Jennifer E.\ Helsby\altaffilmark{2}, Jean-Ren\'e Gauthier\altaffilmark{2,3}, Stephen A.\ Shectman\altaffilmark{3}, Ian B.\ Thompson\altaffilmark{3}, \& Jeremy L.\ Tinker\altaffilmark{4}}

\altaffiltext{1}{This paper includes data gathered with the 6.5 meter Magellan Telescopes located at Las Campanas Observatory, Chile.}
\altaffiltext{2}{Dept.\ of Astronomy \& Astrophysics and Kavli Institute for Cosmological Physics, University of Chicago, Chicago, IL, 60637, U.S.A.
{\tt hchen@oddjob.uchicago.edu}}
\altaffiltext{3}{The Observatories of Carnegie Institution for Science, 813 Santa Barbara St., Pasadena, CA 91101}
\altaffiltext{4}{Berkeley Center for Cosmological Physics, University of California-Berkeley}

\begin{abstract}

  We report results from a survey of Mg\,II absorbers in the spectra
  of background QSOs that are within close angular distances to a
  foreground galaxy at $z<0.5$, using the Magellan Echellette
  Spectrograph.  We have established a spectroscopic sample of 94
  galaxies at a median redshift of $\langle z\rangle = 0.24$ in fields
  around 70 distant background QSOs ($z_{\rm QSO}>0.6$), 71 of which
  are in an 'isolated' environment with no known companions and
  located at $\rho\apll 120\ h^{-1}$ kpc from the line of sight of a
  background QSO.  The rest-frame absolute $B$-band magnitudes span a
  range from $M_{B}-5\log\,h=-16.4$ to $M_{B}-5\log\,h=-21.4$ and
  rest-frame $B_{AB}-R_{AB}$ colors range from $B_{AB}-R_{AB}\approx
  0$ to $B_{AB}-R_{AB}\approx 1.5$.  Of these 'isolated' galaxies, we
  find that 47 have corresponding Mg\,II absorbers in the spectra of
  background QSOs and rest-frame absorption equivalent width
  $W_r(2796)=0.1-2.34$ \AA, and 24 do not give rise to Mg\,II
  absorption to sensitive upper limits.  Our analysis shows that (1)
  \ewr\ declines with increasing distance from 'isolated' galaxies but
  shows no clear trend in 'group' environments; (2) more luminous
  galaxies possess more extended Mg\,II absorbing halos with the
  gaseous radius scaled by $B$-band luminosity according to $R_{\rm
  gas}=75\times (L_B/L_{B_*})^{(0.35\pm 0.03)}\ h^{-1}$ kpc; (3) there
  is little dependence between the observed absorber strength and
  galaxy intrinsic colors; and (4) within $R_{\rm gas}$, we find a
  mean covering fraction of $\langle\kappa_{0.3}\rangle\approx 70$\%
  for absorbers of $\ewr\ge 0.3$ \AA\ and
  $\langle\kappa_{0.1}\rangle\approx 80$\% for absorbers of $\ewr\ge
  0.1$ \AA.  The results confirm that extended Mg\,II absorbing halos
  are a common and generic feature around ordinary galaxies and that
  the gaseous radius is a fixed fraction of the dark matter halo
  radius.  The lack of correlation between \ewr\ strength and galaxy
  colors suggests a lack of physical connection between the origin of
  extended Mg\,II halos and recent star formation history of the
  galaxies.  Finally, we discuss the total gas mass in galactic halos
  as traced by Mg\,II absorbers.  We also compare our results with
  previous studies.

\end{abstract}


\section{INTRODUCTION}

The 'forest' of absorption-line systems observed in the spectra of
background quasars offers a sensitive probe of otherwise invisible
gaseous structures in the universe, where the majority of baryons
reside.  Depending on the physical conditions of the absorbing gas,
different transitions are expected to display different strengths and
kinematic signatures.  Combining QSO absorption-line observations and
faint galaxy surveys, in principle, provides a unique and powerful
means of establishing a comprehensive picture for understanding the
growth of galaxies.

The Mg\,II $\lambda\lambda\,2796, 2803$ doublets are among the
absorption features commonly seen in the spectra of distant quasars.
These absorbers are understood to originate in photo-ionized gas of
temperature $T \sim 10^4$ K (Bergeron \& Stas\'inska 1986; Charlton et
al.\ 2003) and trace high-column density clouds of neutral hydrogen
column density $N(\hI) \approx 10^{18}-10^{22}$ \cmjj\ (Rao et al.\
2006).  At redshifts $z=0.4-2.5$, these doublet features are
redshifted into the optical spectral range and are routinely detected
in the spectra of background QSOs using optical spectrographs on the
ground.  Over the past two decades, a large number of studies have
been carried out to characterize the statistical properties of Mg\,II
absorbers, including the frequency distribution function, redshift
evolution of the absorber number density, and kinematic signatures
(e.g.\ Lanzetta \etal\ 1987; Sargent \etal\ 1988; Petitjean \&
Bergeron 1990; Steidel \& Sargent 1992; Charlton \& Churchill 1998;
Churchill \etal\ 2000, 2003; Nestor \etal\ 2005; Prochter \etal\
2006).  While the accuracy and precision of these various measurements
increase with increasing sample size, the utility of known absorber
statistics in advancing our understanding of galaxy evolution has been
limited due to an ambiguous origin of these absorbers.

A necessary first step toward the goal of applying QSO absorption-line
systems for probing the growth of baryonic structures is to understand
and quantify their correlation with galaxies.  The large associated
\hI\ column density (Rao et al.\ 2006) suggests that Mg\,II absorbers
are similar to those H\,I clouds seen around individual galaxies in
21~cm surveys (e.g.\ Doyle et al.\ 2005).  A direct connection of
these absorbers to galaxies is also supported by the presence of
luminous galaxies at projected distances $\rho = 50-100 \ h^{-1}$ kpc
from known Mg\,II absorbers (Bergeron 1986; Lanzetta \& Bowen 1990,
1992; Steidel \etal\ 1994; Zibetti \etal\ 2005; Nestor \etal\ 2007;
Kacprzak \etal\ 2007).  However, uncertainties remain, because some
galaxies found at $\rho<50\ h^{-1}$ kpc from a QSO sightline do not
produce a corresponding Mg\,II absorber (e.g.\ Tripp \& Bowen 2005;
Churchill \etal\ 2007), implying that Mg\,II absorbers may not probe a
representative population of galaxies or simply that the gas covering
fraction is not unity.

Over the past two years, we have been conducting a program that
combines empirical observations and a phenomenological model study to
establish a comprehensive description of the correlation between
galaxies and cool gas ($T\sim 10^4$ K) probed by Mg\,II absorbers.  In
Tinker \& Chen (2008), we have introduced a novel technique that
adopts the halo occupation framework (e.g.\ Seljak 2000; Scoccimarro
et al.\ 2001; Berlind \& Weinberg 2002) and characterizes the origin
of QSO absorption-line systems based on a conditional mass function of
dark matter halos in which the absorbers are found.  
%The halo
%occupation distribution (HOD) method has been developed to establish
%an empirical mapping between galaxies and dark matter halos, defined
%as collapsed objects with a mean enclosed density of 200 times the
%mean background density (e.g.\ Seljak 2000; Scoccimarro et al.\ 2001;
%Berlind \& Weinberg 2002).  
This technique is purely statistical in nature.  It characterizes the
cold and warm-hot gas in dark matter halos by comparing the frequency
distribution function and clustering amplitude of QSO absorbers with
those of dark matter halos.

Our initial halo occupation model is constrained by known statistical
properties of Mg\,II absorbers at $\langle z\rangle=0.6$, including an
isothermal density profile for describing the spatial distribution of
cold gas in individual dark matter halos, the frequency distribution
function (e.g.\ Steidel \& Sargent 1992; Nestor \etal\ 2005; Prochter
\etal\ 2006), and the clustering amplitude (Bouch\'e \etal\ 2006;
Gauthier \etal\ 2009; Lundgren \etal\ 2009).  The adopted isothermal
model is supported by empirical data (Chen \& Tinker 2008).  The
product of the halo occupation analysis is an occupation function (or
``mass function'') that characterizes the fractional contribution of
dark matter halos (galaxies) of different masses to the observed
Mg\,II absorbers of different strength.

Our analysis has shown that in order to reproduce the observed overall
strong clustering of the absorbers, roughly $5-10$ \% of the gas in
halos up to $10^{14}$ \hmsol\ is required to be cold.  The inferred
presence of cool gas in massive halos is also supported by (i) the
observed strong cross-correlation amplitude of Mg\,II absorbers and
luminous red galaxies (LRG) on projected distance scales of $r_p <
300\ h^{-1}$ comoving kpc (Gauthier \etal\ 2009) and by (ii) direct
detections of Mg\,II absorbers at $\apll 300\ h^{-1}$ kpc and $\apll
320$ \kms\ from five LRGs (Gauthier \etal\ 2010).  These LRGs are
understood to reside in $>10^{13}$ \hmsol\ halos (e.g.\ Blake \etal\
2008; Gauthier \etal\ 2009).  For lower-mass halos, our halo
occupation analysis has shown that the incidence and covering fraction
of extended cool gas is high.  Therefore these halos contribute
significantly to the observed Mg\,II statistics.  The large gas
covering fraction is consistent with the empirical findings of Steidel
\etal\ (1994) and Chen \& Tinker (2008), though other authors have
reported a lower covering fraction from different surveys (e.g.\ Tripp
\& Bowen 2005; Kacprzak \etal\ 2008; Barton \& Cooke 2009).  In
summary, the initial results of our halo occupation analysis
demonstrate that combining galaxy and absorber survey data together
with a simple semi-analytic model already produces unique empirical
constraints for contemporary theoretical models that study the gas
content of dark matter halos (e.g.\ Mo \& Miralda-Escud\'e 1996;
Birnboim \& Dekel 2003; Maller \& Bullock 2004; Kere$\check{\rm s}$
\etal\ 2005; 2009; Dekel \& Birnboim 2006; Birnboim \etal\ 2007).

In Tinker \& Chen (2010), we have expanded upon our initial halo
occupation analysis to incorporate the observed number density
evolution of the absorbers (Nestor \etal\ 2005; Prochter \etal\ 2006)
as an additional constraint to gain insight into the redshift
evolution of extended gas around galaxies.  In order to incorporate
the expected redshift evolution of the dark matter halo population to
explain the observed number density evolution of Mg\,II absorbers, we
have found that the gaseous halos must evolve with respect to their
host dark matter halos.  An explicit prediction of our halo occupation
model is a more pronounced inverse-correlation between the mean halo
mass and absorber strength (e.g.\ Bouch\'e \etal\ 2006; Gauthier
\etal\ 2009; Lundgren \etal\ 2009) at $z\apll 0.3$ and a positive
correlation between the mean halo mass and absorber strength at $z\apg
2$.  However, no clustering measurements are available for Mg\,II
absorbers at $z<0.4$ or $z\apg 1$.

To test this prediction would require a large sample of galaxies and
absorbers from these epochs.  At $z\apg 2$, a large sample of Mg\,II
absorbers is already available (e.g.\ Nestor \etal\ 2005; Prochter
\etal\ 2006) from searches in the Sloan Digital Sky Survey (SDSS; York
\etal\ 2000) quasar sample (e.g.\ Schneider \etal\ 2007), but galaxy
surveys that cover a cosmological volume are challenging in this
redshift range.  In contrast, few Mg\,II absorbers are known at
$z<0.35$ where exquisite details of the galaxy population have been
recorded (e.g.\ the SDSS), due to a lack of spectral sensitivity at
wavelength $\lambda<4000$ \AA\ where the low-redshift Mg\,II
absorption features occur.

A new observing window has become available with recently commissioned
UV sensitive spectrographs on the ground.  In particular, the Magellan
Echellette Spectrograph (MagE; Marshall \etal\ 2008) offers high
throughput over a contiguous spectral range from $\lambda=3100$ \AA\
to 1 $\mu$m, allowing searches for Mg\,II absorbers at redshift as low
as $z=0.11$ (see also Barton \& Cooke 2009 for a similar effort at
$z=0.1$).  Building upon the existing SDSS galaxy database, we have
initiated a MagE survey of Mg\,II absorbers in the spectra of
background QSOs, whose sightlines intercept the halo of a foreground
galaxy at $z\le 0.5$.  The primary goal of this project is to
establish a statistically significant sample ($N\apg 500$) of
low-redshift Mg\,II absorbers for measuring the clustering amplitude
of Mg\,II absorbers at $\langle z\rangle=0.2$ based on the observed
cross-correlation amplitude on co-moving scales of $1-30\ h^{-1}$ Mpc.
Combining the detections and non-detections in the vicinity (impact
separation of $\rho\apll 100\ h^{-1}$ physical kpc) of known galaxies
will also facilitate a comprehensive study of how the properties of
extended cool gas (such as the density profile and covering fraction)
correlate with known stellar and ISM properties of the host galaxies.
Here we introduce the MagE Mg\,II absorber survey and present initial
results from the first year of data.

This paper is organized as follows.  In Section 2, we describe the
design of our MagE survey project.  In Section 3, we describe the
spectroscopic observations of photometrically selected galaxies in the
SDSS data archive and the MagE follow-up of quasars selected from the
SDSS spectroscopic QSO catalog.  We provide a summary of the data
reduction and analysis procedures.  In Section 4, we present the
catalogs of galaxies and Mg\,II absorbers.  In Section 5, we examine
the correlation between Mg\,II absorption strength and galaxy
properties.  Finally, we discuss the properties of extended cool gas
in galactic halos and compare our results with previous studies in
Section 6.  We adopt a $\Lambda$CDM cosmology, $\Omega_{\rm M}=0.3$
and $\Omega_\Lambda = 0.7$, with a dimensionless Hubble constant $h =
H_0/(100 \ {\rm km} \ {\rm s}^{-1}\ {\rm Mpc}^{-1})$ throughout the
paper.

\section{EXPERIMENT DESIGN}

Previous surveys of Mg\,II absorbers have been carried out at $z>
0.35$ (e.g.\ Steidel \& Sargent 1992; Nestor \etal\ 2005; Prochter
\etal\ 2006), where the Mg\,II $\lambda\lambda\,2796, 2803$ doublet
features are redshifted into the optical spectral window at wavelength
$\lambda > 3800$ \AA.  Few absorbers are known at lower redshifts
(Jannuzi \etal\ 1998; Bechtold \etal\ 2002).  At the same time,
wide-field surveys such as the Sloan Digital Sky Survey (SDSS; York
\etal\ 2000) have yielded detailed maps of the large-scale galaxy
distribution over $1/4$ of the sky.  Nearly $10^8$ galaxies brighter
than $r'=22$ are identified at $z<0.5$ in the SDSS archive using
photometric redshift techniques.  Photometric redshifts determined
from the observed broad-band spectral discontinuities are found to be
accurate to within an r.m.s. residual of $|z_{\rm phot}-z_{\rm
  spec}|=0.03$ for galaxies of $r'<20$ and $|z_{\rm phot}-z_{\rm
  spec}|=0.13$ for galaxies of $r'>20$ (e.g.\ Oyaizu \etal\ 2008).
These photometrically selected galaxies, when combined with QSO
absorbers along common sightlines, provide a unique opportunity for a
comprehensive study of the physical origin of the absorbers and for
characterizing the properties of extended gas around galaxies.

We use MagE, which offers high throughput over a contiguous spectral
range from $\lambda=3100$ \AA\ to 1 $\mu$m, to conduct a survey of
Mg\,II absorbers at $z=0.11-0.35$.  We aim to establish the first
statistically significant sample of Mg\,II absorbers ($N\apg 500$) at
$z<0.4$, using MagE on the Magellan Clay Telescope.  The primary
objectives of our survey are (1) to examine the cool gas content in
halos around galaxies based on the presence/absence of Mg\,II
absorption features; (2) to study the correlation between Mg\,II
absorption strength and known galaxy properties such as luminosity,
impact distance, and stellar population; and (3) to quantify the
origin of Mg\,II absorbers at $z \sim 0.2$ based on their large-scale
clustering properties.  The results, when combined with known
statistical properties of Mg\,II absorbers at $z=0.4-1$, will allow us
to establish an empirical characterization of the evolution of cool
baryons in dark matter halos (e.g.\ Tinker \& Chen 2010).

To maximize the efficiency of the Mg\,II absorber search, we select
those QSO sightlines that are close to a foreground galaxy with
photometric redshift $z_{\rm phot}\le 0.4$.  The maximum angular
separation for the selected QSO--galaxy pairs is determined based on
previous observations of extended Mg\,II absorbing gas around
intermediate-redshift galaxies.  Using 23 galaxy--Mg\,II absorber
pairs at $z=0.207-0.892$, Chen \& Tinker (2008) found a distinct
boundary at $\rho={\hat R}_{\rm gas}$, beyond which no Mg\,II
absorbers with rest-frame absorption equivalent width $\ewr>0.01$ \AA\
are found.  At $\rho<{\hat R}_{\rm gas}$, the incidence of Mg\,II
absorbers was found to be $\approx 80-86$\%.  In addition, the
observed correlation between $\ewr$ and the physical projected
distance $\rho$ to the galaxy is well described by an isothermal
density profile with ${\hat R}_{\rm gas}$ scaled by galaxy $B$-band
luminosity ($L_B$) or absolute $B$-band magnitude ($M_B$) according to
\begin{equation}
\frac{\hat R_{\rm gas}}{\hat R_{\rm gas*}}=\left (\frac{L_B}{L_{B_*}}\right )^{0.35}=10^{0.14\,(\,M_B\,-\,M_{B_*})},
\end{equation}
and
\begin{equation}
{\hat R}_{\rm gas*}=91\ h^{-1}\,{\rm kpc}.
\end{equation}

We select galaxies from the SDSS Data Release 6 (DR6) archive, and
adopt Equations (1) and (2) as the fiducial model for calculating the
expected gaseous radii for galaxies in the SDSS Data Release 6 (DR6)
archive.  The rest-frame absolute $B$-band magnitude of each galaxy is
estimated using the best-fit photometric redshift $z_{\rm phot}$ from
Oyaizu \etal\ (2008)\footnote{We have adopted the photometric
redshifts evaluated under case 'cc2' in Oyaizu \etal\ (2008), which
included four colors $u'-g'$, $g'-r'$, $r'-i'$, $i'-z'$, and three
concentration indices in the $g'r'i'$ bands.  According to these
authors, 'cc2' provides the best and most realistic redshift
estimates.} and its model $g'$-band magnitude from the SDSS
archive\footnote{All object magnitudes quoted in this paper are in the
$AB$ system..  To remind the readers, the SDSS imaging survey reaches
a 5-$\sigma$ limiting magnitude of $AB=22$ mag in the $u'$, $g'$, and
$r'$ bands, and $AB\approx 21$ mag in the $i'$ and $z'$ bands (York
\etal\ 2000).  In selecting our galaxies, we do not require a
detection in all five bands.  At a limiting magnitude of $r'=22$, we
will be able to uncover $L_*$ galaxies at $z=0.4$ and sub-$L_*$
galaxies at lower redshifts.}, following
\begin{equation}
M_B = g' - 5\,\log\,\frac{D_L}{\rm 10\,pc} + 2.5\,\log\,(1+z) - k(z),
\end{equation}
where $k(z)$ is the $k$-correction term to account for the color
difference between the observed $g'$ band and corresponding rest-frame
$B$ band.  We estimate the $k$-correction of each galaxy using an
$S_{\rm bc}$ galaxy template from Coleman \etal\ (1980), and adopt
$M_{B_*}-5\,\log\,h=-19.8$ from Faber \etal\ (2007) when computing
$\hat{R}_{\rm gas}$ using Equations (1) and (2).

From a cross-comparison between the SDSS DR6 catalogs of galaxies and
QSOs (Adelman-McCarthy et al.\ 2008), we identify QSO and galaxy pairs
according to the following criteria.  First, we consider galaxies at
$z_{\rm phot}\le 0.4$ to explore the unique spectral window offered by
MagE for the Mg\,II absorber search.  We consider QSOs that are
brighter than $u'=19.5$ in order to obtain sufficient $S/N$ in a
relatively short exposure using MagE on the Magellan Clay Telescope.
We select QSOs at $z\le 1.5$ to avoid contamination due to the \lya\
forest.  Next, we impose additional constraints to select only QSOs at
$z>0.4$ and at more than 10,000 \kms\ behind the galaxy in the
redshift space, in order to minimize the incidence of correlated QSO
and galaxy pairs (e.g.\ Wild \etal\ 2008).  Finally, we select
QSO--galaxy pairs with projected separations less than ${\hat R}_{\rm
gas}$ from Equations (1) and (2).  This procedure yields $\approx
3900$ close QSO--galaxy pairs at Declination $< +20^{\circ}$ with
angular separation $\theta < 180''$ on the sky.

The galaxies in the pair sample span a broad range in intrinsic
colors, luminosity, and impact separation.  Although they are
pre-selected to occur within the fiducial radius of $\rho={\hat
R}_{\rm gas}$, uncertainties in the photometric redshifts allow us to
study the gaseous halo in regions beyond ${\hat R}_{\rm gas}$.

\section{Observations and Data Reduction}
 
Our program requires spectroscopic data of both galaxies and absorbers
along common lines of sight toward background quasars.  Here we
describe the observation and data reduction procedures.

\subsection{Galaxy Spectroscopy}

We have obtained optical spectra of 89 photometrically selected SDSS
galaxies, using the Double Imaging Spectrograph (DIS; Lupton \etal\
1995) on the 3.5~m telescope at the Apache Point Observatory and MagE
(Marshall \etal\ 2008) on the Magellan Clay Telescope at the Las
Campanas Observatory.  All of these galaxies satisfy the selection
criteria described in \S\ 2.  The medium-to-high resolution spectra
allow us to obtain precise and accurate redshift measurements of these
galaxies, which are necessary for establishing (or otherwise) a
physical connection with absorbers found along nearby QSO sightlines.
The spectroscopic observations of these SDSS galaxies are summarized
here.

Long-slit spectra of 41 galaxies were obtained using DIS over the
period from October 2008 through June 2009.  The blue and red cameras
have a pixel scale of $0.4''$ and $0.42''$ per pixel, respectively, on
the 3.5~m telescope.  We used a $1.5''$ slit, the B400 grating in the
blue channel for a dispersion of 1.83 \AA\ per pixel, and the R300
grating in the red channel for a dispersion of 2.31 \AA\ per pixel.
The blue and red channels with the medium-resolution gratings together
offer contiguous spectral coverage from $\lambda=3800$ \AA\ to
$\lambda=9800$ \AA\ and a spectral resolution of ${\rm FWHM} \approx
500$ \kms\ at $\lambda=4400$ \AA\ and ${\rm FWHM} \approx 400$ \kms\
at $\lambda=7500$ \AA.  The observations were carried out in a series
of two to three exposures of between 1200 s and 1800 s each, and no
dither was applied between individual exposures.  Flat-field frames
were taken in the afternoon prior to the beginning of each night.
Calibration frames for wavelength solutions were taken immediately
after each set of science exposures using the truss lamps on the
secondary cage.  When the sky was clear, we also observed a
spectrophotometric standard star observed at the beginning of each
night for relative flux calibration.

All of the DIS spectroscopic data were reduced using standard
long-slit spectral reduction procedures.  The spectra were calibrated
to vacuum wavelengths, corrected for the heliocentric motion, and
flux-calibrated using a sensitivity function derived from observations
of a flux standard.  Redshifts of the galaxies were determined based
on a cross-correlation analysis with a linear combination of SDSS
eigen spectra of galaxies. The typical redshift uncertainty was found
to be $\Delta z \sim 0.0003$.  An example of the DIS galaxy spectra is
presented in the top panel of Figure 1.

\begin{figure}
\begin{center}
\includegraphics[scale=0.3,angle=270]{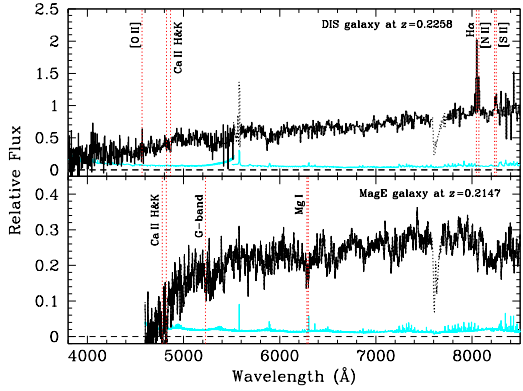}
\caption{{\it Top}: An example DIS spectrum of a galaxy
  (SDSSJ130555.49$+$014928.62) at $z=0.2258$.  {\it Bottom}: An
  example MagE spectrum of a galaxy (SDSSJ230845.53$-$091445.97) at
  $z=0.2147$.  Contaminating sky features are dotted out. The
  corresponding 1-$\sigma$ error spectrum is shown at the bottom of
  each panel.}
\end{center} 
\label{gal_spec}
\end{figure}

Optical spectra of 48 galaxies were obtained using MagE in September
2008, February 2009, March 2009, and June 2009.  The camera has a
plate scale of $0.3''$ per pixel.  We used a $1''$ slit and $2\times
1$ binning during readout, which yielded a spectral resolution of
${\rm FWHM} \approx 150$ \kms.  Depending on the brightness of the
galaxies, the observations were carried out in a sequence of one to
two exposures of duration 300 s to 1200 s each.  The galaxy data were
processed and reduced using data reduction software that was
originally developed by G.\ Becker and later modified by us to work
with binned spectral frames.  Wavelengths were calibrated using a ThAr
frame obtained immediately after each exposure and subsequently
corrected to vacuum and heliocentric wavelengths.  Flux calibration
was performed using a sensitivity function derived from observations
of the flux standard GD50.  Individual flux-calibrated orders were
coadded to form a single spectrum that covers a spectral range from
$\lambda\approx 4000$ \AA\ to $\lambda=9600$ \AA.  Redshifts of the
galaxies were determined based on a cross-correlation analysis with a
linear combination of SDSS eigen spectra of galaxies. The typical
redshift uncertainty was found to be $\Delta z \sim 0.0001$.  An
example of the MagE galaxy spectra is presented in the bottom panel of
Figure 1.

We have also located five additional galaxies, which satisfy the
selection criteria described in \S\ 2 and already have optical spectra
and robust spectroscopic redshifts available in the public SDSS data
archive.  These galaxies are included in our final galaxy sample for
the search of Mg\,II absorbers.  A journal of the spectroscopic
observations of the full galaxy sample is in Table 1.

\begin{figure}
\begin{center}
\includegraphics[scale=0.3,angle=270]{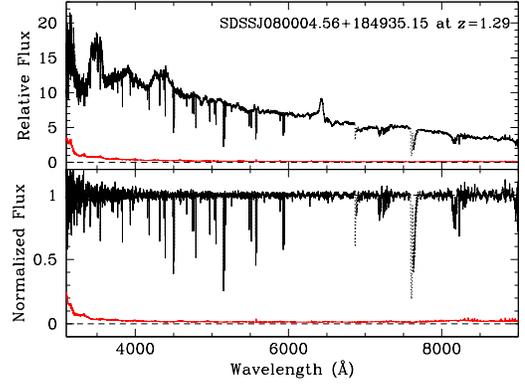}
\caption{{\it Top}: An order-combined MagE spectrum of
  SDSSJ080004.56$+$184935.15 at $z=1.29$.  {\it Bottom}: Continuum
  normalized, stacked spectrum of the same QSO.  Absorption features
  due to the Atmosphere are dotted out.  This sightline displays
  multiple strong Mg\,II absorbers at $z=0.254-1.12$, including one at
  $z=0.254$ at $\rho=22.2\ h^{-1}$ kpc and $\Delta\,v=-191$ \kms\ from
  a known galaxy.  The corresponding 1-$\sigma$ error spectrum is
  shown at the bottom of each panel.}
\end{center} 
\label{mage_qso}
\end{figure}

\subsection{Echellette Spectra of QSOs}

\begin{figure*}
\begin{center}
\includegraphics[scale=0.3]{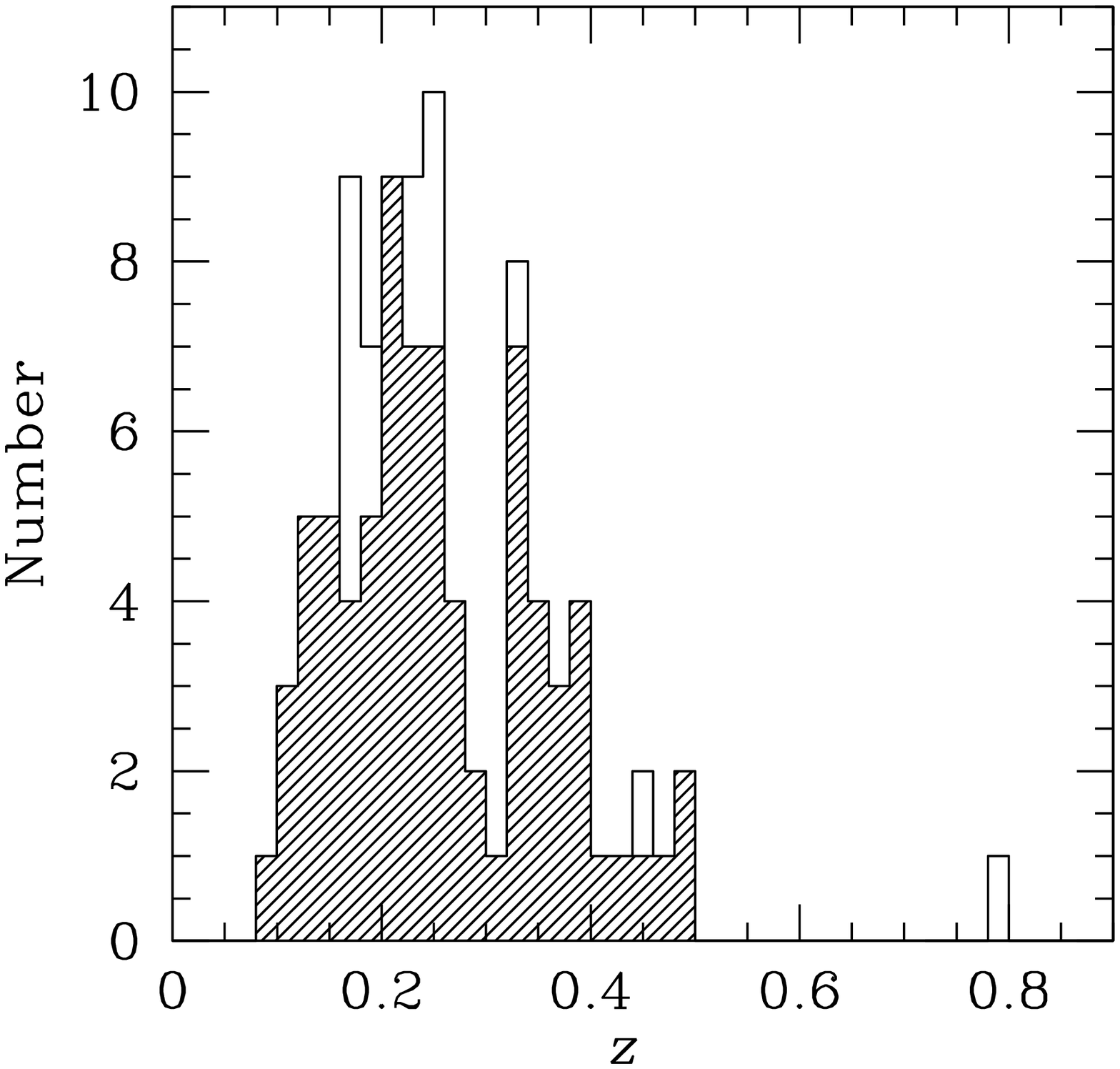}
\includegraphics[scale=0.3]{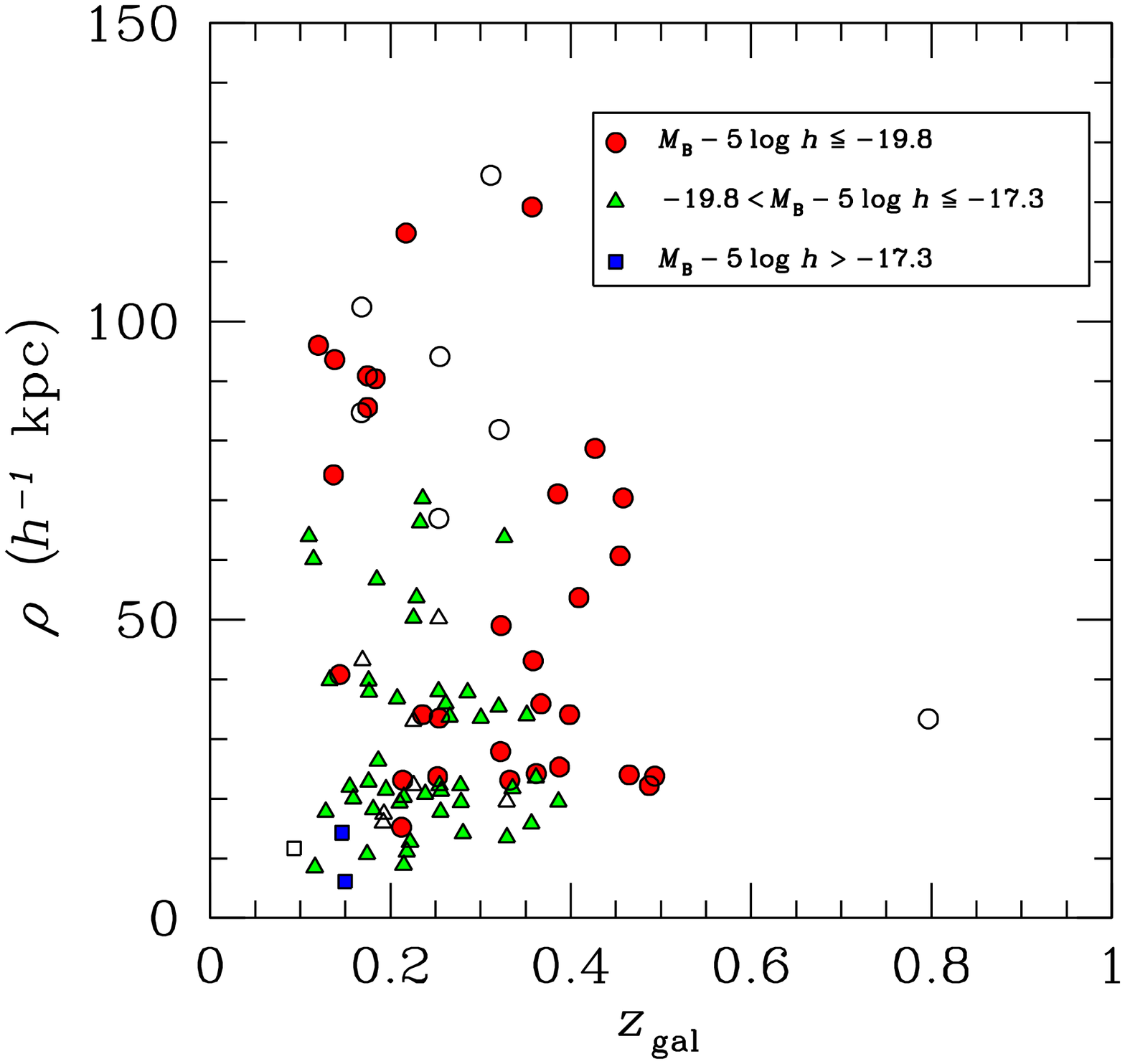}
\caption{{\it Left}: Redshift distritbutions of 94 galaxies in the
  'full' sample (solid histogram) and of 77 galaxies in the 'isolated'
  galaxy sample (shaded histogram).  {\it Right}: Distribution of
  projected separations between the spectroscopically confirmed
  galaxies and their nearby QSO sightlines.  Galaxies of $B$-band
  luminosity $L_B>L_{B_*}$ are shown in circles; galaxies of
  $L_B=(0.1-1)\,L_{B_*}$ in triangles; and galaxies of
  $L_B<0.1\,L_{B_*}$ in squares.}
\end{center} 
\label{zgal}
\end{figure*}

Echellette spectroscopic observations of 70 QSOs were obtained using
the MagE spectrograph (Marshall \etal\ 2008) on the Magellan Clay
telescope over the period from January 2008 through June 2009.  We
used a $1''$ slit for the majority of the QSOs (with a few sources
taken with a $0.7''$ slit) and $1\times 1$ binning during readout,
which yielded a spectral resolution of ${\rm FWHM} \approx 70$ \kms\
(or ${\rm FWHM} \approx 50$ \kms\ for spectra taken with a $0.7''$
slit).  Depending on the brightness of the QSOs, the observations were
carried out in a sequence of one to three exposures of duration 600 s
to 2400 s each.  The QSO spectra were processed and reduced using a
reduction pipeline developed by G.\ Becker and kindly offered to us
by the author.  Wavelengths were calibrated using a ThAr frame obtained
immediately after each exposure and subsequently corrected to vacuum
and heliocentric wavelengths.  Flux calibration was performed using a
sensitivity function derived from observations of the flux standard
GD50.  Individual flux calibrated echellette orders were coadded to
form a single spectrum that covers a spectral range from
$\lambda=3050$ \AA\ to $\lambda=1\,\mu$m.  These order-combined
individual exposures were then continuum normalized and stacked to
form one final combined spectrum per QSO using an optimal weighting
routine.  The continuum was determined using a low-order polynomial
fit to spectral regions that are free of strong absorption features.
The flux calibrated spectra have $S/N\apg 10$ per resolution element
across the entire spectral range.  Examples of an order-combined
spectrum and a continuum-normalized final stack are shown in Figure 2.
A journal of the MagE QSO observations is in Table 2.

\section{The Galaxy and Mg\,II Absorber Catalogs}

\subsection{The Galaxy Sample}

We have established a sample of 94 galaxies with $r'<22.3$ and robust
spectroscopic redshifts at $z<0.8$ in fields around 70 distant
background QSOs ($z_{\rm QSO}>0.6$).  These 94 spectroscopically
confirmed galaxies form the 'full' sample of our study.  Within the
full sample, 17 galaxies are found with at least one close neighbor at
$\rho<{\hat R}_{\rm gas}$ and velocity separation $\Delta\,v<300$
\kms, including one galaxy that occurs in the vicinity of a background
QSO at $z\approx 0.8$.  The presence of a close neighbor implies that
the galaxy is likely to reside in a group environment.  The
association between absorbers and individual galaxies becomes
uncertain in a group environment.  In addition, interactions between
group members are expected to alter the properties of gaseous halos
(e.g.\ Gunn \& Gott 1972; Balogh \etal\ 2000; Verdes-Montenegro \etal\
2001).  Therefore, these galaxies form a 'group'-galaxy subsample and
are considered separately from the remaining 'isolated'-galaxy
subsample in our analysis below.  Here we summarize the galaxy
properties in the full sample.

We first examine the redshift distribution of the galaxies.  Figure 3
shows in the left panel the redshift distribution of all 94
spectroscopically confirmed galaxies in our survey (full histogram)
and the redshift distribution of the 'isolated' galaxy subsample
(shaded histogram).  Figure 4 shows the comparison of photometric
redshifts $z_{\rm phot}$ and spectroscopic redshift $z_{\rm spec}$ for
galaxies in our full sample.  Excluding the single outlier at
$z\approx 0.8$, we find that SDSS photometric redshifts for galaxies
brighter than $r'=21.6$ are accurate to within a mean residual of
$\langle\,(z_{\rm phot}-z_{\rm spec})/(1+z_{\rm spec})\,\rangle=0.01$
and a corresponding r.m.s. scatter of
$\delta\,z\equiv\Delta\,z/(1+z)=0.06$.  There exists an increased
uncertainty in the photometric redshifts for galaxies at $z\apg 0.35$,
resulting in the apparent 'tilt' seen in the left panel.  We also
examine photometric redshift uncertainties versus galaxy brightness
and intrinsic color.  The right panel of Figure 4 shows that the
precision and accuracy of photometric redshifts are still higher for
brighter ($r'< 20$) and redder galaxies (with rest-frame
$B_{AB}-R_{AB}$ color consistent with elliptical/S0 galaxies; see
Equation 4 and Figure 5 for the classification of different galaxy
types).  With the exception of one outlier at $z\approx 0.8$, the
target selection based on known photometric redshifts has allowed us
to effectively identify foreground galaxies at $z= 0.1-0.4$ that occur
at close projected distances to the line of sight toward a background
QSO.

\begin{figure*}
\begin{center}
\includegraphics[scale=0.3]{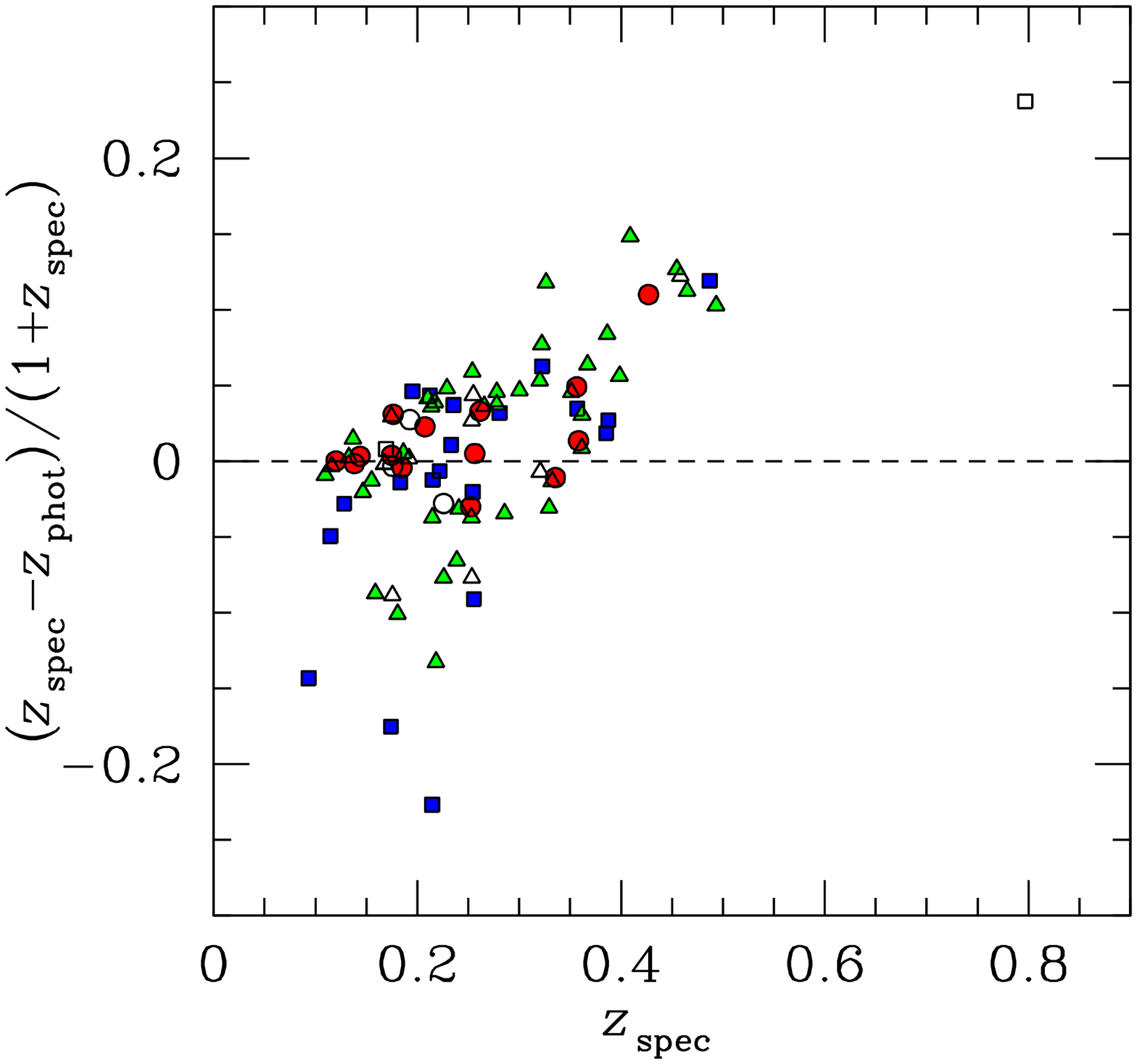}
\includegraphics[scale=0.3]{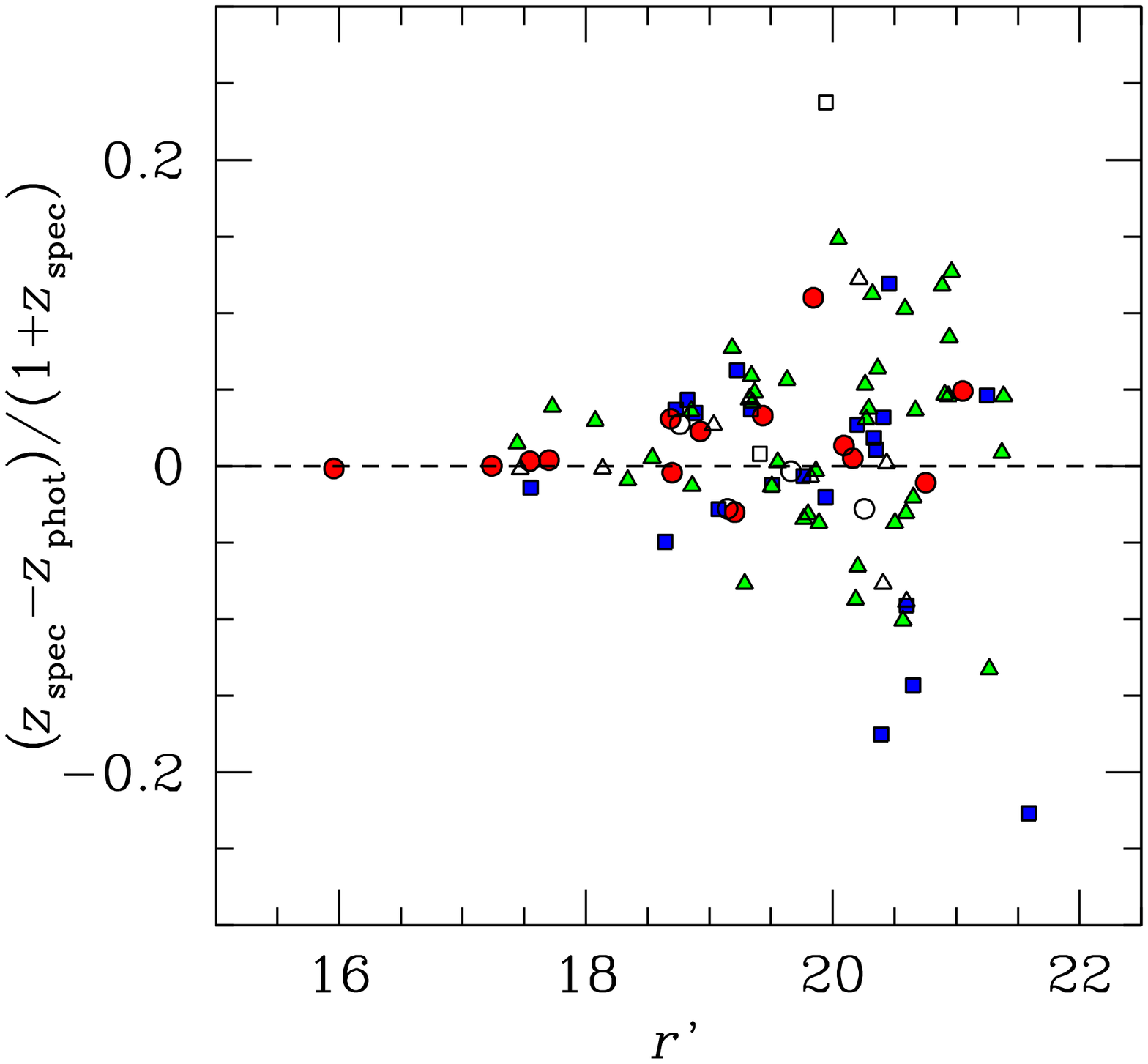}
\caption{Comparisons of photometric redshifts $z_{\rm phot}$ and
  spectroscopic redshift $z_{\rm spec}$ for galaxies at different
  redshift (left panel) and of different $r'$-band magnitude (right
  panel).  Galaxies with intrinsic $B_{AB}-R_{AB}$ color consistent
  with elliptical/S0 galaxies are shown in circles; disk galaxies are
  shown in triangles; and irregular/star-forming galaxies are shown in
  squares (see Equation 4 and Figure 5 for the classification of
  different galaxy types).  Galaxies in the 'group' subsample are
  shown in open symbols.}
\end{center} 
\label{zcomp}
\end{figure*}

Next, we present the impact parameter distribution of the galaxies to
a QSO sightline in the right panel of Figure 3.  'Group' galaxies are
displayed in open symbols and 'isolated' galaxies in solid symbols.
Galaxies in different luminosity ranges are shown in different shapes
of symbols.  Figure 3 shows that luminous galaxies with $B$-band
luminosity $L_B>L_{B_*}$ (circles)\footnote{We adopt
$M_{B_*}-5\,\log\,h=-19.8$ from Faber \etal\ (2007), which best
characterizes the blue galaxy population at $z\sim 0.4$.}  span a
range in their projected distance to a QSO sightline from $\rho=15.2\
h^{-1}$ kpc to $\rho=119.2\ h^{-1}$ kpc, while sub-$L_*$ galaxies of
$L_B=(0.1-1)\,L_{B_*}$ (triangles) cover a range from $\rho=8.5\
h^{-1}$ kpc to $\rho=70.3\ h^{-1}$ kpc.  Our sample also includes two
faint dwarfs with $L_B<0.1\,L_{B_*}$ (squares) at $\rho<15\ h^{-1}$
kpc.  In addition, the QSO point spread function prevents us from
finding galaxies at angular distances $\Delta\,\theta\apll 3''$ from
the QSO sightlines in the SDSS catalog.  The minimum project distances
we can probe using the SDSS galaxy sample changes from $\rho_{\rm
min}<10\ h^{-1}$ kpc at $z\apll 0.15$ to $\rho_{\rm min}>15\ h^{-1}$
kpc at $z\apg 0.4$.  The distribution shows that our survey is
designed to study halo gas (rather than the interstellar medium) of
distant galaxies.  Our sample includes more luminous galaxies
(presumably more massive and therefore residing in more extended
gaseous halos) at larger impact parameters with the maximum survey
radius determined by the fiducial model (\S\ 2).

The rest-frame absolute $B$-band magnitudes and rest-frame $B-R$
colors of all 94 galaxies are calculated based on the spectroscopic
redshifts.  To carry out this calculation, we first adopt the best-fit
model magnitudes from the SDSS photometric catalog and compute the
composite model magnitudes in the $u'g'r'i'z'$ bandpasses following
the instructions of Scranton \etal\ (2005)\footnote{The composite
  model magnitude is evaluated based on the composite flux, $f_{\rm
    composite}$, which is a linear combination of the best-fit fluxes
  contained in an exponential profile $f_{\rm exp}$ and a de
  Vaucouleurs profile $f_{\rm deV}$: $f_{\rm composite} = frac_{\rm
    deV}\,f_{\rm deV}+(1-frac_{\rm deV})\,f_{\rm exp}$.  The composite
  model magnitude is considered to provide the best-estimate of the
  total flux of a galaxy (see Scranton \etal\ 2005 for more detailed
  discussion).}.  Then, the composite model magnitudes are corrected
for the Galactic extinctions following Schlegel \etal\ (1998).
Finally for each galaxy, we identify the bandpasses that correspond
most closely to the rest-frame $B$ and $R$ bands based on its
spectroscopic redshift.  We interpolate between the composite model
magnitudes to determine the intrinsic $B-R$ color and the $k$-term
necessary to correct the remaining color difference for deriving the
rest-frame absolute $B$-band magnitude following Equation (3).

\begin{figure*}
\begin{center}
\includegraphics[scale=0.3]{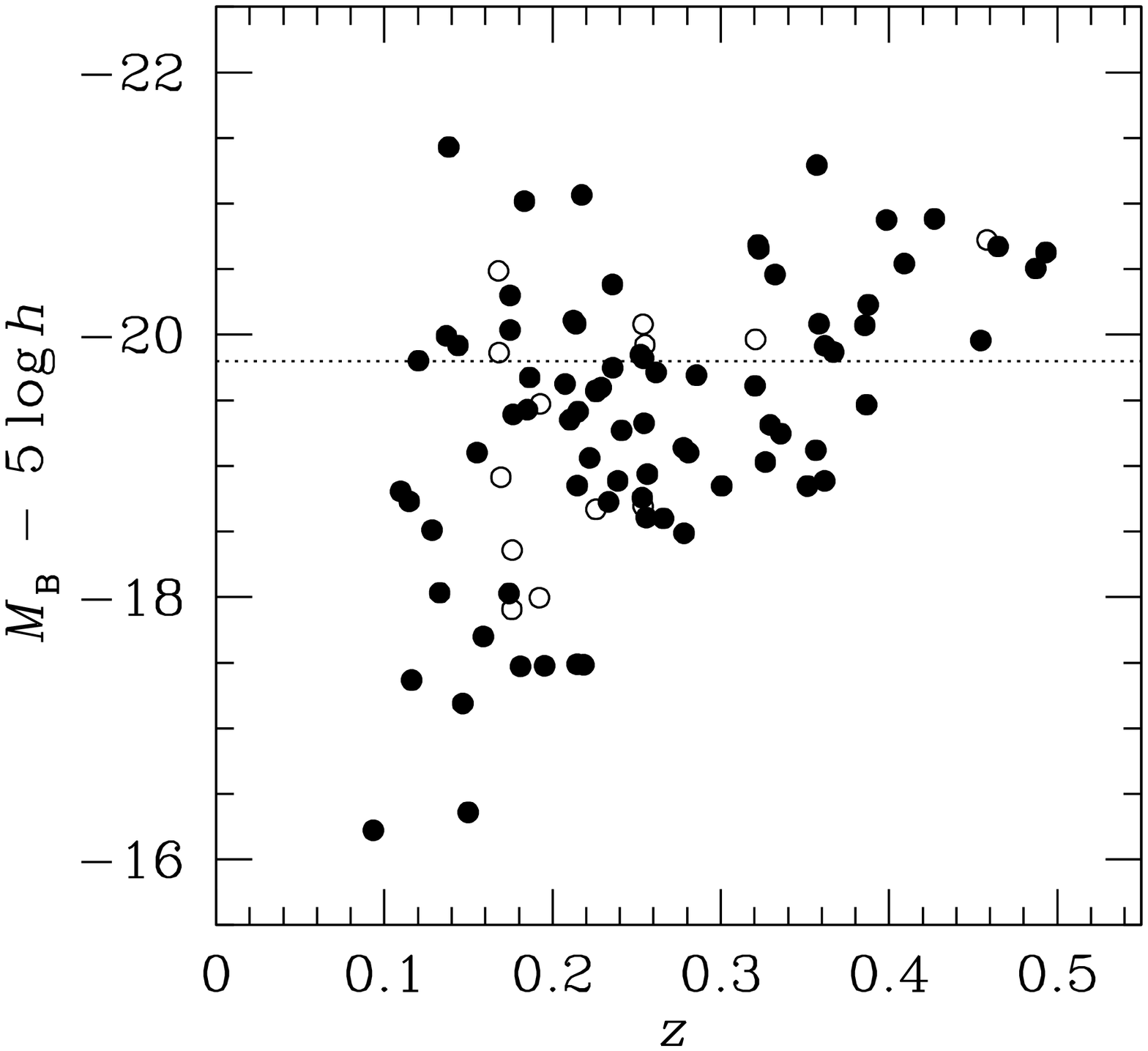}
\includegraphics[scale=0.3]{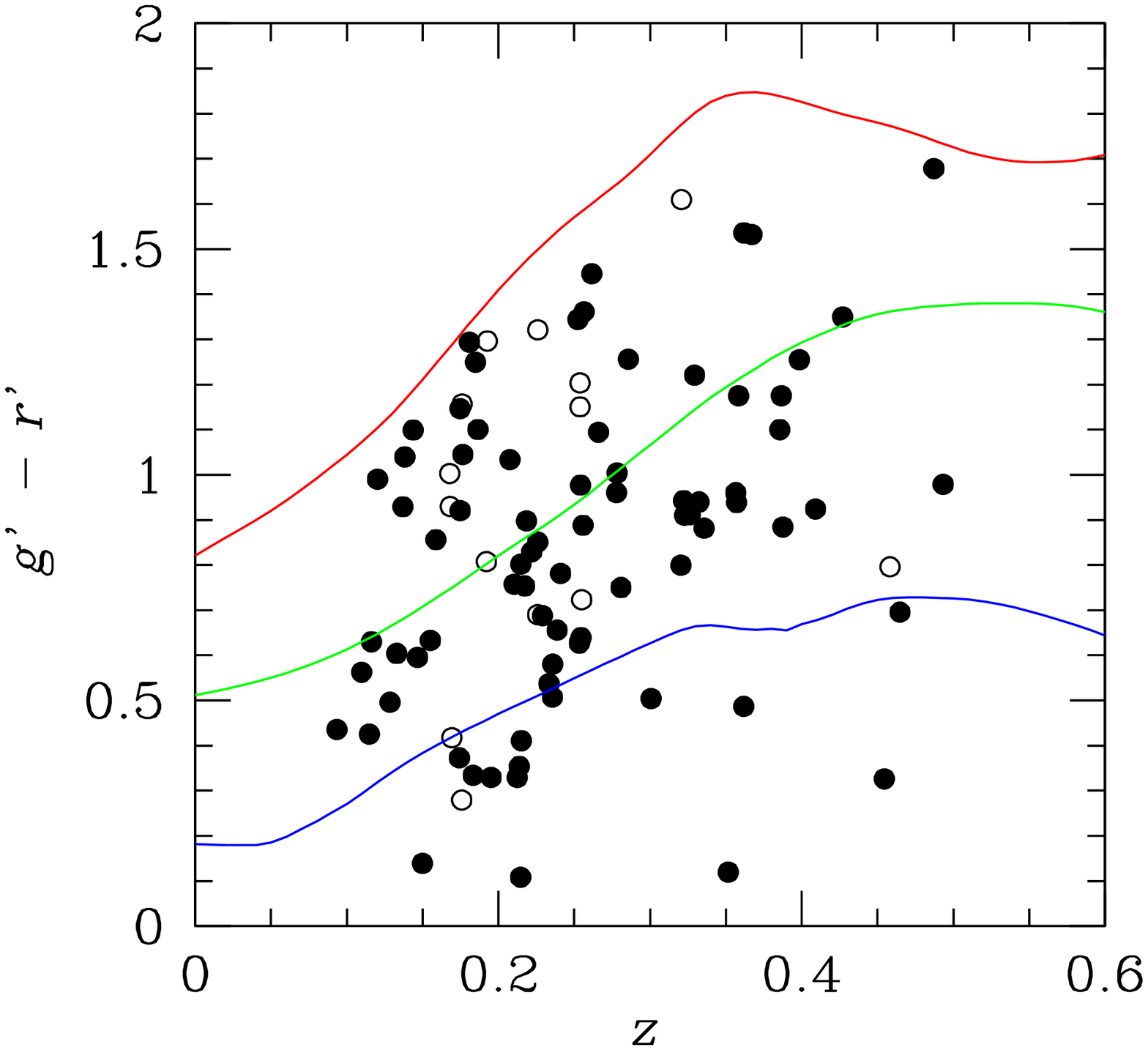}
\caption{Redshift distributions of rest-frame absolute $B$-band
  magnitudes (left) and observed $g'-r'$ colors (right) of the
  spectroscopically confirmed galaxies in the 'group' subsample (open
  histograms/symbols) and in the 'isolated' subsample (shaded
  histograms/closed symbols).  The galaxies span a range in the
  rest-frame absolute $B$-band magnitude from $M_{B}-5\log\,h=-16.22$
  to $M_{B}-5\log\,h=-22.5$, and a broad range in intrinsic colors.
  The dotted line in the left panels marks $M_{B_*}$ from Faber \etal\
  (2007).  The solid curves in the right panel indicate the expected
  $g'-r'$ color versus redshift based on non-evolving galaxy templates
  for E/S0 (top curve), Sbc (middle curve), and irregular (bottom
  curve) galaxies from Coleman \etal\ (1980).}
\end{center} 
\label{mgal}
\end{figure*}

Figure 5 shows the redshift distribution of rest-frame absolute
$B$-band magnitudes of our galaxies, spanning a range from
$M_{B}-5\log\,h=-16.22$ to $M_{B}-5\log\,h=-22.5$.  We also examine
the underlying stellar population by comparing the observed optical
colors with expectations from known spectral energy distribution
templates of galaxies of different morphological type.  We consider
the E/S0, Sbc, Scd, and Irr galaxy templates from Coleman \etal\
(1980) and calculate the expected colors versus redshift under a
no-evolution assumption (see also Fukugita \etal\ 1995).  The right
panel of Figure 5 shows that the optical colors of our galaxies are in
broad agreement with the expected colors of galaxies across a broad
range of morphology.  We classify the galaxies based on the intrinsic
$B_{AB}-R_{AB}$ color according to the following criteria:
\begin{eqnarray}
&&B_{AB}-R_{AB} > 1.1\hspace{0.45in}\mbox{elliptical/S0,} \nonumber\\
&&0.6 < B_{AB}-R_{AB} \le 1.1\hspace{0.1in}\mbox{disk galaxies,} \nonumber \\
&&B_{AB}-R_{AB} < 0.6\hspace{0.45in}\mbox{irregular/star-forming galaxies.}
\end{eqnarray}
Of the 94 galaxies in our spectroscopic sample, 18 have the intrinsic
$B_{AB}-R_{AB}$ color consistent with an elliptical or S0 galaxy, 52
are consistent with a disk galaxy, and 24 are consistent with an
irregular or younger galaxies.  The classification is qualitatively
consistent with the spectral features seen in the galaxy spectra.  Of
the 18 galaxies classified as elliptical or S0, more than $2/3$
display spectral features dominated by absorption transitions due to
Ca\,II H\&K and G-band.  A more detailed analysis is presented in a
separate paper (Helsby \etal\ in preparation).

\subsection{The Galaxy--Mg\,II Absorber Pair Sample}

To establish a galaxy--Mg\,II absorber pair sample for the subsequent
analysis, we examine the echellette spectra of the background QSOs and
search for the corresponding Mg\,II absorption doublet at the
locations of the galaxies presented in \S\ 4.1.  Specifically, we
first search for the corresponding doublet features within velocity
separation $\Delta\,v=\pm\,1000$ \kms\ of the galaxy redshifts.  Then
we accept absorption lines according to a $2\,\sigma$ detection
threshold criterion, which is appropriate because the measurements are
performed at known galaxy redshifts.  Next, we determine the absorber
redshift based on the best-fit line centroid of a Gaussian profile
analysis of the doublets.  Uncertainties in the absorber redshifts are
less than $\delta\,v=(c\,\Delta\,z/(1+z)=25$ \kms.  Next, we measure
2-$\sigma$ upper limits to the 2796-\AA\ absorption equivalent widths
for galaxies that are not paired with corresponding absorbers.

\begin{figure*}
\begin{center}
\includegraphics[scale=0.5]{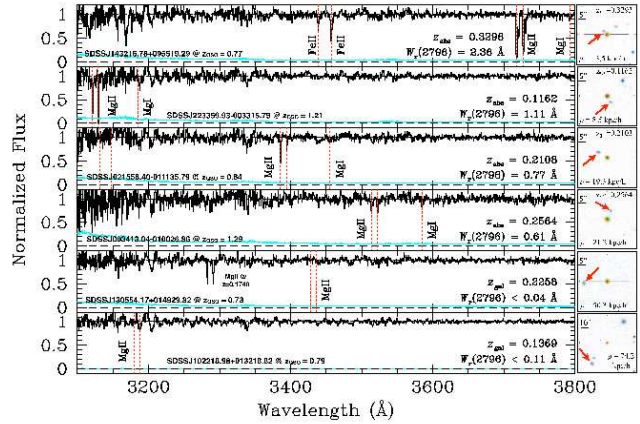}
\caption{Examples of the galaxy--Mg\,II absorber pairs in our sample.
For each field, we present the QSO spectrum in the left panel and the
corresponding SDSS image centered at the QSO in the right panel.  The
galaxies in our survey with confirmed spectroscopic redshifts is
indicated by the arrow.  The projected distance to the QSO is shown in
the lower right corner of each image panel.  The absorption features
corresponding to the redshifts of the galaxies are indicated by dotted
lines in the left panels.  The absorber redshift is shown in the lower
right corner of each spectrum.  For non-detections in the bottom two
panels, we note the redshift of the galaxy and the 2-$\sigma$ upper
limit of \ewr.}
\end{center} 
\end{figure*}

The procedure identifies 47 Mg\,II absorbers and 24 upper limits in
the vicinities of 71 isolated galaxies, and seven Mg\,II absorbers and
one upper limit around 17 'group' galaxies.  We are unable to obtain
significant constraints for the presence/absence of Mg\,II absorbers
around five galaxies, because the expected Mg\,II absorption features
are blended with either strong C\,IV $\lambda\lambda\,1548,1550$
absorption features at higher redshifts or the atmosphere O\,3
absorption complex at $\lambda \approx 3200$.  Finally, one galaxy is
found at $z_{\rm spec}=0.0934$, outside of the redshift range targeted
for the MagE Mg\,II absorber survey.  Examples of the galaxy--Mg\,II
absorber pairs, including upper limits, are presented in Figure 6.  A
summary of the properties of the spectroscopically confirmed galaxies
is presented in Table 3, where we list for each galaxy in columns
(1)---(8) the ID, Right Ascension and Declination offsets from the QSO
$\Delta \alpha$ and $\Delta \delta$, redshift $z_{\rm gal}$, impact
parameter $\rho$, apparent $r'$-band magnitude, $B_{AB}-R_{AB}$ color,
and absolute $B$-band magnitude $M_B - 5 \log h$.  Measurement
uncertainties in $r'$ and $M_B - 5 \log h$ are typically $0.05$ dex.

Impact parameter separations of the galaxy and absorber pairs in the
'isolated' sample range from $\rho = 6.1$ to $119.2 \ h^{-1}$ physical
kpc with a median of $\langle \rho\rangle_{\rm med} = 27.9 \ h^{-1}$
kpc.  The redshifts of the 'isolated' galaxies range from $z = 0.1097$
to $z=0.4933$ with a median of $\langle z\rangle_{\rm med} = 0.2533$.
The rest-frame absolute $B$-band magnitudes of the galaxies span a
range from $M_{B}-5\log\,h=-16.4$ to $M_{B}-5\log\,h=-21.4$ with a
median of $\langle M_{B}-5\log\,h\rangle=-19.6$.  The rest-frame
$B_{AB}-R_{AB}$ colors range from $B_{AB}-R_{AB}\approx 0$ to
$B_{AB}-R_{AB}\approx 1.5$ with a median of $\langle
B_{AB}-R_{AB}\rangle_{\rm med}\approx 0.8$. 

\begin{figure}
\begin{center}
\includegraphics[scale=0.4]{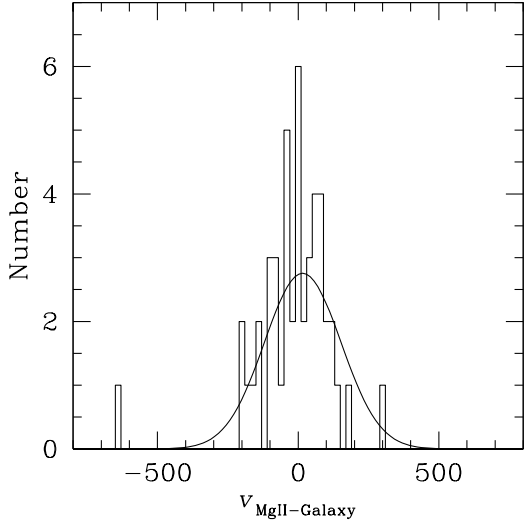}
\caption{Relative velocity Distribution of the 47 galaxy--Mg\,II
  absorber pairs in the 'isolated' sample.  The pair sample is well
  characterized by a Gaussian distribution (the solid curve) with a
  mean velocity difference of $\langle\,v_{\tiny\rm
    Mg\,II-Galaxy}\rangle=16$ \kms\ and dispersion of $\sigma=137$
  \kms. }
\end{center} 
\end{figure}

The redshift and absorption equivalent width of the corresponding
Mg\,II absorption feature for each of the 71 galaxies in our sample
are listed in columns (9)---(10) of Table 3.  Figure 7 shows the
relative velocity distribution of the 47 galaxy--Mg\,II absorber pairs
in the isolated sample.  We find that with the exception of one pair,
which exhibits a velocity separation of $\,v_{\tiny\rm
Mg\,II-Galaxy}=-645$ \kms, the velocity separation between the Mg\,II
absorbers and their member galaxies is well characterized by a
Gaussian distribution of mean velocity difference
$\langle\,v_{\tiny\rm Mg\,II-Galaxy}\rangle=16$ \kms\ and dispersion
$\sigma=137$ \kms.  The velocity dispersion between the galaxy--Mg\,II
absorber pair sample is comparable to the velocity dispersion seen in
a Milky-Way size galaxy (see also Steidel \etal\ 2002; Chen \etal\
2005; Kacprzak \etal\ 2010 for rotation curve comparisons of galaxies
and strong absorbers at intermediate redshifts), supporting a physical
association between the absorbing gas and the member galaxy.

\section{Analysis}

We examine the correlation between galaxies and Mg\,II absorbers
identified along common lines of sight, using the sample of 47
galaxy--Mg\,II absorber pairs and 24 galaxies at $\rho\apll 120\
h^{-1}$ kpc that do not give rise to Mg\,II absorption to a sensitive
upper limit.  Figure 8 shows the distribution of \ewr versus $\rho$
for our 'isolated' sample of 71 galaxies.  Galaxies with intrinsic
$B_{AB}-R_{AB}$ color consistent with elliptical/S0 galaxies are shown
in circles, those consistent with disk galaxies are shown in
triangles; and irregular/star-forming in squares.  Points with arrows
in Figure 8 indicate $2 \sigma$ upper limits.  Galaxies in the 'group'
subsample are also included as open symbols for comparison.  Similar
to previous results (e.g.\ Lanzetta \& Bowen 1990; Kacprzak \etal\
2008), we find that on average \ewr\ declines with increasing $\rho$
for galaxies in an 'isolated' environment.  In contrast, while
galaxies from a 'group' environment appear to occupy a similar
$W_r(2796)$ versus $\rho$ space with 'isolated' galaxies, they do not
exhibit a strong inverse correlation.  In this section, we describe
the methods we use for obtaining and assessing the best-fit models
that represent the data and present the results.

\begin{figure}
\begin{center}
\includegraphics[scale=0.4]{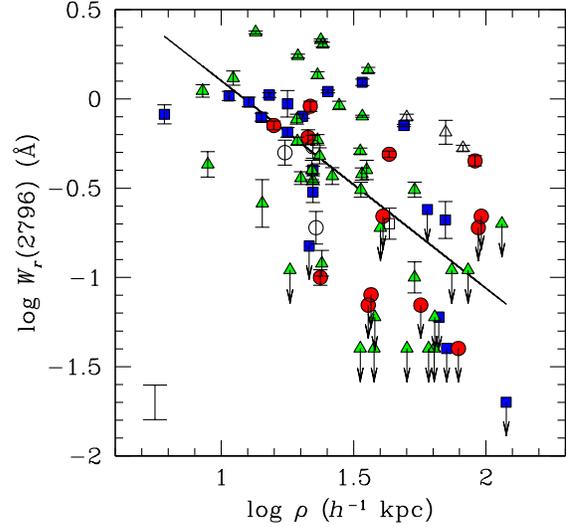}
\caption{Comparison of the corresponding rest-frame absorption
  equivalent width $W_r(2796)$ versus galaxy impact parameter $\rho$
  for a sample of 71 galaxies.  Circles represent galaxies with
  intrinsic $B_{AB}-R_{AB}$ colors consistent with elliptical or S0
  galaxies, triangles represent early- and late-type disk galaxies,
  and squares represent irregular, star-forming galaxies.  Points with
  arrows indicate $2 \sigma$ upper limits. The solid line indicate the
  best-fit power-law model.  The errorbar in the lower-left corner
  indicates the intrinsic scatter estimated based on the likelihood
  analysis discussed in \S\ 5.1 and Equation (7).  We also include
  galaxies that are in the 'group' subsample (open symbols) for
  comparisons.}
\end{center} 
\end{figure}

\subsection{Method}

  To obtain the best-fit model that characterizes the correlation
between the strength of Mg\,II absorbers and the properties of
galaxies, we adopt a parameterized functional form
\begin{equation}
y = f(x1,x2,..),
\end{equation}
where $y$ is the predicted absorber strength and is always
$\log\,W_r(2796)$ in our analysis, and $x_i$'s are independent
measurements of galaxy properties such as galaxy impact parameter
$\rho$, rest-frame absolute magnitude $M_B-5\,\log\,h$, redshift $z$,
or intrinsic color $B_{AB}-R_{AB}$.  We consider a simple power-law
model for characterizing the \ewr\ versus $\rho$ relation.  In
logarithmic space, the model is expressed as a linear function
according to
\begin{equation}
\log\,\bar{W}_r^{\rm p}(2796)\,=\,a_0 + a_1 \log\,\rho + a_2 (M_B-M_{B_*}) + ....
\end{equation}

To determine the values of different coefficients $a_i$'s, we perform
a maximum-likelihood analysis that includes the upper limits in our
galaxy and Mg\,II absorber pair sample.  The likelihood function of
this analysis is defined as
\begin{eqnarray}
{\cal L}(\bar{y}) = & &\left( \prod_{i=1}^{n} \exp \left\{ -\frac{1}{2} \left[ \frac{y_i -
\bar{y}(\rho_i, L_{B_i})}{\sigma_i} \right]^2 \right\} \right) \times \nonumber \\ 
& & \left(\prod_{i=1}^m \int_{y_i}^{-\infty} dy' \exp \left\{ -\frac{1}{2} \left[ 
\frac{y' - \bar{y}(\rho_{i}, L_{B_i})}{\sigma_i} \right]^2 \right\} \right),
\end{eqnarray}
where $y_i=\log\,W_i$ is the observed value of galaxy $i$,
$\bar{y}=\log\,\bar{W}_r(2796)$ is the model expectation, and
$\sigma_i$ is the measurement uncertainty of $y_i$.  The first product
of Equation (7) extends over the $n$ measurements and the second
product extends over the $m$ upper limits.  (This definition of the
likelihood function is appropriate if the residuals about the mean
relationship are normally distributed.)  We also consider the
possibility that $\sigma_i$ includes a significant intrinsic scatter
(which presumably arises due to intrinsic variations between
individual galaxies) as well as measurement error.  We express
$\sigma_i$ as a quadratic sum of the cosmic scatter $\sigma_c$ and the
measurement error $\sigma_{m_i}$
\begin{equation}
\sigma_i^2 = \sigma_c^2 + \sigma_{m_i}^2,
\end{equation}
where the intrinsic scatter is defined by
\begin{equation}
\sigma_c^2 =  {\rm med} \left( \left\{ y_i - \bar{y}(\rho_{i},L_{B_i},...) -
\frac{1}{n} \sum_{j=1}^n \left[ y_j - \bar{y}(\rho_{j},L_{B_j},...) \right]
\right\}^2 - \sigma_{m_i}^2 \right).
\end{equation}
Because $\sigma_c$ depends on the maximum-likelihood solution
$\bar{y}=\log\,\bar{W}(\rho_{i},L_{B_i})$, the maximum-likelihood
solution is obtained iteratively with respect to equations (7) and
(9).

\subsection{Dependence of Extended Gas on Galaxy Impact Parameter and $B$-band Luminosity}

Adopting a power-law function for describing the $W_r(2796)$ versus
$\rho$ distribution in Figure 8, we find based on the
maximum-likelihood analysis a best-fit model of
\begin{equation}
\log \bar{W}_r^{\rm p}(2796) = -(1.17\pm 0.10) \log \rho + (1.28\pm 0.13)
\end{equation}
with an intrinsic scatter of $\sigma_c=0.195$.  The r.m.s.\ residual
between the observed and model Mg\,II absorber strengths is found to
be ${\rm r.m.s.}(\log\,W_r-\log\,\bar{W})=0.350$.  Errors associated
with the best-fit coefficients are 1-$\sigma$ uncertainties.  We
present the best-fit model and the intrinsic scatter in Figure 8.

The results of the likelihood analysis demonstrates at a high
confidence level that the Mg\,II absorber strength $W_r(2796)$ scales
inversely with galaxy impact parameter.  To assess the significance of
this anti-correlation, we perform a generalized Kendall test
(Feigelson \& Nelson 1985) that accounts for the presence of
non-detections and find that the observed distribution of $W_r(2796)$
versus $\rho$ deviates from a random distribution at more than
5.5-$\sigma$ level of significance.

Despite a statistically significant anti-correlation seen between
$W_r(2796)$ and $\rho$, there exists a large scatter between the data
and the best-fit power-law model (Equation 10).  Previous authors
(Chen \& Tinker 2008; Kacprzak \etal\ 2008) have shown that the
observed Mg\,II absorber strength scales with galaxy $B$-band
luminosity.  Recall that our galaxy sample spans a range in rest-frame
absolute $B$-band magnitude from $M_{B}-5\log\,h=-16.4$ to
$M_{B}-5\log\,h=-21.4$.  Here we examine whether including intrinsic
$B$-band luminosity helps to further strengthen this anti-correlation.

Adopting a power-law function to parameterize the scaling relation, we
find based on the likelihood analysis that the observed absorber
strength is best described by $\log W_r(2796) = -(1.52\pm 0.10) \log
\rho - (0.20\pm 0.03)\,(M_B-M_{B_*}) + (1.88\pm 0.15)$ (dotted line in
Figure 9), with an intrinsic scatter of $\sigma_c=0.175$ and ${\rm
  r.m.s.}(\log\,W_r(2796)-\log\,\hat{W})=0.318$.  We adopt
$M_{B_*}-5\,\log\,h=-19.8$ from Faber \etal\ (2007), which best
characterizes the blue galaxy population at $z\sim 0.4$.  The result
of the likelihood analysis shows that including the scaling with
$B$-band luminosity indeed reduces the intrinsic scatter by $\approx
12$\%.  At the same time, five outliers at $>3\,\sigma_c$ from the
mean relation (points in dotted circles) have now become apparent
after accounting for the scaling with $B$-band luminosity.

Inspecting the imaging and spectral properties of the outliers, we
note that the discrepant triangle and circle in the lower-center part
of the graph are, respectively, an active galaxy showing broad
emission lines and a red, evolved galaxy with a possible companion
blended with a nearby bright star.  The discrepant triangle in the
upper-left is the galaxy-Mg\,II absorber pair with a large velocity
separation of $\Delta\,v\approx -645$ \kms\ in Figure 7.  The
discrepant triangle and circle in the upper-right are galaxies with
likely companions at close distances.  Follow-up spectroscopy of these
surrounding galaxies will confirm the nature of these close neighbors.

\begin{figure}
\begin{center}
\includegraphics[scale=0.4]{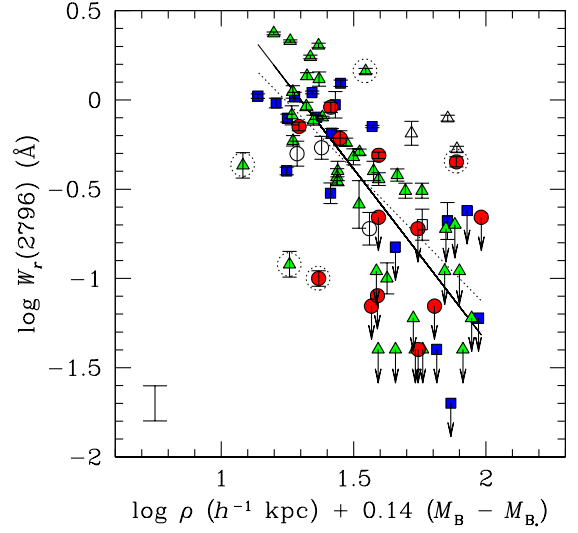}
\caption{Comparison of $W_r(2796)$ versus $B$-band luminosity scaled
  impact parameter $\log\,\rho'=\log\,\rho+a_2'\,(M_B-M_{B_*})$.  The
  scaling coefficient $a_2'\equiv a_2/a_1$ (Equation 6) is determined
  based on the likelihood analysis described in \S\ 5.1 and is found
  to be $a_2'=0.14\pm 0.01$.  Symbols are the same as in Figure 8. The
  dotted line is the best-fit power-law model for the entire sample of
  71 galaxies, while the solid line is the best-fit power-law model,
  excluding five outliers (according to a 3-$\sigma$ clipping
  criterion; see \S\ 5.2) marked in dotted circles.  We note that the
  discrepant triangle and circle in the lower center part of the graph
  are, respectively, an active galaxy showing broad emission lines and a
  red, evolved galaxy with a possible companion blended with a nearby
  bright star.  The discrepant triangle in the upper-left region is
  the galaxy-Mg\,II absorber pair with a large velocity separation of
  $\Delta\,v\approx -645$ \kms\ in Figure 7.  The errorbar in the
  lower-left corner indicates the intrinsic scatter estimated from the
  likelihood analysis and defined in Equation (7).  }
\end{center} 
\end{figure}

Excluding the five outliers, we obtain a revised best-fit power-law
model of
\begin{eqnarray}
\log \bar{W}_r^{\rm p}(2796) &=& -(1.93\pm 0.11) \log \rho \nonumber \\ 
                            & & - (0.27\pm 0.02)\,(M_B-M_{B_*}) \nonumber \\ 
                            & & + (2.51\pm 0.16),
\end{eqnarray}
with an intrinsic scatter of $\sigma_c=0.196$ and ${\rm
  r.m.s.}(\log\,W_r-\log\,\bar{W})=0.235$.  We find that after
accounting for an optimal scaling relation with $L_B$, the slope of
the anti-correlation is steepened.  We present the revised best-fit
model as solid line in Figure 9.  A generalized Kendall test
demonstrates that the distribution of $W_r(2796)$ versus $\rho$ after
accounting for galaxy luminosity deviates from a random distribution
at more than 7-$\sigma$ level of significance.  It indicates that the
probability that $W_r(2796)$ is randomly distributed with respect to
$\rho$ is $\ll 0.1$\%.

\subsection{Dependence of Extended Gas on Galaxy Color and Redshift}

Here we examine whether including additional galaxy properties, such
as $B_{AB}-R_{AB}$ color and redshift, helps to further strengthen
this anti-correlation.  Recall that our galaxy sample spans a range in
$B_{AB}-R_{AB}$ color from $B_{AB}-R_{AB}\approx 0$ to
$B_{AB}-R_{AB}\approx 1.5$, and a range in redshift from $z=0.1097$ to
$z=0.4933$.

Including the intrinsic $B_{AB}-R_{AB}$ color of the galaxies as an
additional independent variable in the power-law model and excluding
the outliers in Figure 9, we find based on the likelihood analysis
(Equations 7 through 9) that
\begin{eqnarray}
\log \bar{W}_r^{\rm p}(2796) &=& -(1.92\pm 0.11) \log \rho \nonumber \\ 
                            & & - (0.27\pm 0.02)\,(M_B-M_{B_*}) \nonumber \\ 
                            & & + (0.05\pm 0.02)\,(B_{AB}-R_{AB}) \nonumber \\
                            & & + (2.46\pm 0.15),
\end{eqnarray}
with an intrinsic scatter of $\sigma_c=0.184$ and ${\rm
  r.m.s.}(\log\,W_r-\log\,\bar{W})=0.246$.  The small scaling
coefficient of the color term in Equation (12) indicates that the
observed $W_r(2796)$ versus $\rho$ anti-correlation depends only
weakly on galaxy intrinsic colors.

Next, we examine whether including galaxy redshift as an additional
independent variable in the power-law model improves the strength of
the anti-correlation.  Excluding the outliers, we obtain a best-fit
model of
\begin{eqnarray}
\log \bar{W}_r(2796) &=& -(1.90\pm 0.11) \log \rho \nonumber \\ 
                     & & - (0.25\pm 0.03)\,(M_B-M_{B_*}) \nonumber \\ 
                     & & + (0.73\pm 0.50)\,\log \frac{(1+z)}{(1+z_0)} \nonumber \\ 
                     & & + (2.47\pm 0.15),
\end{eqnarray}
where $z_0=0.25$ which is the median redshift of the 'isolated' galaxy
sample.  The best-fit relation has an intrinsic scatter of
$\sigma_c=0.184$ and ${\rm r.m.s.}(\log\,W_r-\log\,\bar{W})=0.245$.
The uncertain scaling coefficient of the redshift term in Equation
(13) indicates that there is little evolution in the extended Mg\,II
halos around galaxies between $z\approx 0.1$ and $z\approx 0.5$.

\section{DISCUSSION}

We have established a spectroscopic sample of 94 galaxies at a median
redshift of $\langle z\rangle = 0.2357$ in fields around 70 distant
background QSOs ($z_{\rm QSO}>0.6$).  Our follow-up MagE spectra of
the QSOs allow us to examine the extent of cold gas around 88 of these
galaxies based on the presence/absence of coincident Mg\,II absorption
features.  

In the sample of 88 spectroscopically confirmed galaxies, 17 are found
in a 'group' environment with two or more close neighbors at
$\rho<{\hat R}_{\rm gas}$ (Equation 1) and velocity separation
$\Delta\,v<300$ \kms.  Excluding one galaxy that occurs in the
vicinity of the background QSO, we identify seven galaxy 'groups'
along the lines of sight of the background QSOs in our sample.  All
seven galaxy 'groups' have coincident Mg\,II absorbers.  Because the
association between absorbers and individual galaxies becomes
uncertain in a group environment, we have separated these group
galaxies from the remaining 'isolated' ones.  Figure 12 shows that
while galaxies from a 'group' environment appear to occupy a similar
range in the $W_r(2796)$ versus $\rho$ parameter space with 'isolated'
galaxies, no strong inverse correlation is seen in the gaseous
profiles of 'group' galaxies.  Because interactions between group
members are expected to alter the properties of gaseous halos, such as
ram pressure and tidal stripping that could re-distribute cold gas to
larger radii (e.g.\ Gunn \& Gott 1972; Balogh \etal\ 2000;
Verdes-Montenegro \etal\ 2001), the 'group'-galaxy subsample also
presents a unique opportunity to study gas kinematics in overdense
galaxy environments beyond the local universe (c.f.\ Verdes-Montenegro
\etal\ 2001).
%Detailed studies of the morphologies and kinematics of
%group galaxies are expected to provide additional insights into the
%physical origin of these Mg\,II absorbers in group environments.

The remaining 71 'isolated' galaxies are located at $\rho\apll 120\
h^{-1}$ kpc from the line of sight of a background QSO.  We identify
47 coincident Mg\,II absorbers in the QSO spectra with absorber
strengths varying from $W_r(2796)=0.1$ \AA\ to $W_r(2796)=2.34$ \AA,
and measure a sensitive upper limit of the Mg\,II absorber strength
for the remaining 24 galaxies.  In the absence of a complete
spectroscopic survey of galaxies around the 71 'isolated' galaxies, it
is likely that some fraction of these galaxies also occur in a group
environment.  However, the strong $\ewr$ versus $\rho$
anti-correlation after accounting for the luminosity scaling relation
seen for the 'isolated' galaxies in Figures 9 \& 12 indicates that
with the exception of a few outliers the majority of the 'isolated'
galaxies are indeed different from those 'group' galaxies.  We
therefore argue that the majority of the 'isolated' galaxies are in a
more quiescent environment than the 'group' ones in our sample.

The results of our likelihood analysis demonstrates that the Mg\,II
absorber strength $W_r(2796)$ scales inversely with galaxy impact
parameter.  In addition, the $W_r(2796)$ vs.\ $\rho$ anti-correlation
is still stronger after accounting for the scaling relation with
galaxy $B$-band luminosity, indicating that more luminous galaxies are
surrounded by more extended Mg\,II halos.  However, including
intrinsic $B_{AB}-R_{AB}$ color does not improve the observed
$W_r(2796)$ vs.\ $\rho$ anti-correlation, indicating a lack of
physical connection between the origin of extended Mg\,II halos and
recent star formation history of the galaxies (c.f.\ Zibetti \etal\
2007).  Finally, the $W_r(2796)$ vs.\ $\rho$ anti-correlation appears
to depend only weakly on galaxy redshift, indicating little evolution
of the extended Mg\,II halos between $z\approx 0.5$ and $z\approx
0.1$.  Here we focus on the 'isolated' galaxy sample and discuss the
covering fraction and spatial profile of extended Mg\,II absorbing gas
based on the observations and analysis presented in \S\ 5.  We also
discuss implications for the cool baryon content of galactic halos and
compare our results with previous studies.

\subsection{Incidence and Covering Fraction of Mg\,II Absorbers}

The analysis presented in \S\ 5.2 shows that the strengths of Mg\,II
absorbers depend on the projected distances and intrinsic luminosities
of the absorbing galaxies.  In addition to the best-fit scaling
relation, we also note that with the exception of two outliers all 33
galaxies at luminosity-scaled impact parameter $\rho'\equiv
\rho\times\,(L_B/L_{B_*})^{-0.35}<30\ h^{-1}$ (Equation 11) have a
coincident Mg\,II absorber of $W_r(2796)>0.3$ \AA.  The observed high
incidence of strong Mg\,II absorbers around galaxies with a broad
range of intrinsic colors strongly supports the notion that extended
Mg\,II halos are a common and generic feature of galaxies of all
types, from evolved early-type galaxies to late-type star-forming
systems.

While extended Mg\,II absorbing gas reaches out to larger projected
distances, our sample shows that both the absorption strength and gas
covering fraction declines toward larger radii.
%Given that our experiment is designed to
%identify a mean statistical trend between the properties of Mg\,II
%absorbers and the properties of galaxies based on an ensemble of
%galaxy and absorber pairs, we cannot distinguish between a partial gas
%covering fraction and a lack of extended gaseous halos.  Here we
%examine the integrated quantity, the product of the incidence of
%%gaseous halos and the covering fraction of Mg\,II absorbing gas.  We
%adopt a single parameter $\kappa$ to represent this integrated
%quantity and 
Here we examine how the gas covering fraction $\kappa$ varies with
$\rho$ and $W_r(2796)$.

We perform a maximum-likelihood analysis to determine $\kappa$.  The
probability that a galaxy gives rise to an absorption system of some
absorption equivalent width threshold is written as
\begin{equation}
P(\kappa)=\kappa(\rho,W_0) \, B [r_1(L_B),r_2(L_B);\rho],
\end{equation}
where $\kappa$ is the fraction of galaxies that give rise to Mg\,II
absorption, and $B$ is a boxcar function that defines the impact
parameter interval; $B=1$ if $r_1(L_B)\le \rho < r_2(L_B)$ and $B=0$
otherwise.  Equation (14) takes into account the scaling relation
between the gaseous extent of galaxies and galaxy $B$-band luminosity.
The likelihood of detecting an ensemble of galaxies, $n$ of which give
rise to Mg\,II absorption systems and $m$ of which do not, is given by
\begin{eqnarray}
{\cal L}(\kappa)&=&\prod_{i=1}^n\,\kappa(\rho_i,W_i) B[r_1(L_{B_i}),r_2(L_{B_i});\rho_i] \nonumber \\
                & & \times
\prod_{j=1}^m\,\{1-\kappa(\rho_j,W_j) B[r_1(L_{B_j}),r_2(L_{B_j});\rho_j]\} \nonumber \\
        &=& \prod_{i=1}^n\,\langle\kappa\rangle\,\prod_{j=1}^m(1-\langle\kappa\rangle) \nonumber \\
        &=& \langle\kappa\rangle^n\,(1-\langle\kappa\rangle)^m.
\end{eqnarray}

We evaluate $\kappa$ for three absorption equivalent width thresholds,
$W_0=0.5$ \AA, $W_0=0.3$ \AA, and $W_0=0.1$ \AA.  In the 'isolated'
galaxy sample, all MagE spectra of the corresponding QSOs have
sufficient $S/N$ for detecting an absorber of $W_r(2796)\ge 0.3$ \AA\
and 61 have sufficient $S/N$ spectra for uncovering an absorber of
$W_r(2796)\ge 0.1$ \AA.  We calculate $\kappa$ for different impact
parameter intervals, using the best-fit scaling relation of Equation
(11) and excluding the outliers.  The results are shown in Figure 10.
The errorbars indicate the 68\% confidence interval of the estimated
$\kappa(\rho,W_0)$.  The impact parameter intervals are chosen to
contain ten galaxy--absorber pairs per bin, except for the last bin
that includes only the remaining seven pairs at largest separations.
Figure 10 shows that the covering fraction of $W_r(2796)\ge 0.1$ \AA\
absorbers varies from 100\% within $\rho=30\ h^{-1}$ kpc of an $L_*$
galaxy to $<20$\% beyond $\rho=60\ h^{-1}$ kpc.

\begin{figure}
\begin{center}
\includegraphics[scale=0.4]{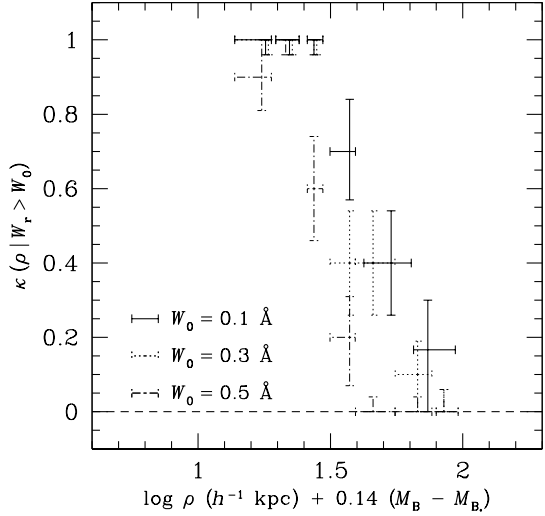}
\caption{Incidence and covering fraction of Mg\,II absorbers versus
  galaxy impact parameter accounting for the galaxy $B$-band
  luminosity scaling relation (Equation 11).  Solid points indicate
  absorbers of $W_r(2796)\ge 0.1$ \AA; dotted points indicate
  absorbers of $W_r(2796)\ge 0.3$ \AA; and dot-dashed points indicate
  absorbers of $W_r(2796)\ge 0.5$ \AA.  The impact parameter intervals
  are chosen to contain ten galaxy--absorber pairs per bin, except for
  the last bin that includes only the remaining seven pairs at largest
  separations.  The errorbars represent the 68\% confidence interval.}
\end{center} 
\end{figure}

The large gas covering fraction seen in our galaxy sample at small
projected distances is in stark contrast to the finding of Gauthier
\etal\ (2010), who have compared the incidence of Mg\,II absorbers
around luminous red galaxies (LRGs) of $i'<20$ mag at $z\approx 0.5$.
These authors have found that the covering fraction of extended Mg\,II
absorbers of $W_r(2796)\ge 0.5$ \AA\ is {\em no more than} $40$\% at
$\rho<50\ h^{-1}$ kpc (or $\rho'<35\ h^{-1}$ kpc after accounting for
the luminosity scaling) and $<30$\% at $\rho<100\ h^{-1}$ kpc (or
$\rho'<70\ h^{-1}$ kpc) from LRGs.  At $W_r(2796)\ge 0.5$, our galaxy
sample shows $\kappa_{0.5}\approx 83$\% at $\rho'<30\ h^{-1}$ kpc.

We note that the LRGs are luminous with $M_B-5\,\log\,h<-21.35$ and
are understood to reside in massive halos of $M_h \apg 10^{13}
\hmsol$, whereas the galaxies in our sample are fainter, with a median
rest-frame $B$-band magnitude of $\langle
M_{B}-5\log\,h\rangle=-19.6$, and presumably reside in lower-mass
halos of $M_h\sim 10^{12.3} \hmsol$ at $z\approx 0.25$.  To examine
how the covering fraction of Mg\,II absorbing gas varies with galaxy
luminosity, we first divide our galaxy sample into luminous
($L_B>L_{B_*}$) and faint ($L_B\le L_{B_*}$) subsamples and then
determine the mean gas covering fraction $\langle\kappa_{W_0}\rangle$
over the entire gaseous halo defined by $R_{\rm gas}$ (see \S\ 6.2) at
a given \ewr\ threshold, $W_0$.

We consider three different threshold values, $W_0=0.1$ \AA, $W_0=0.3$
\AA, and $W_0=0.5$ \AA.  The results are presented in Figure 11,
together with the measurement of Gauthier \etal\ (2010) for $W_0=0.5$
\AA\ around LRGs.  The decreasing mean covering fraction over a fixed
halo radius with increasing absorber strength is consistent with the
expectation from a decreasing cloud density profile toward large radii
(see \S\ 6.2).  In addition, we find little dependence between the
mean gas covering fraction and galaxy luminosity within our sample,
but a significant reduction in gas covering fraction around LRGs.  The
declining gas covering fraction of strong Mg\,II absorbers with
increasing galaxy luminosity (halo mass) is qualitatively consistent
with the expectation from the observed clustering properties of Mg\,II
absorbers (Tinker \& Chen 2008, 2009) and the theoretical expectation
of diminishing cool gas fraction in massive halos (Kere\v{s} \etal\
2005, 2009; Dekel \& Birnboim 2006).

\begin{figure}
\begin{center}
\includegraphics[scale=0.4]{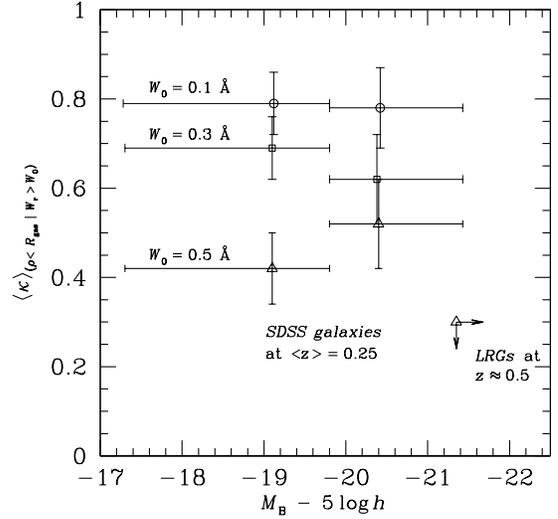}
\caption{Mean covering fraction of Mg\,II absorbers within $R_{\rm
    gas}$ of galaxies in different luminosity intervals.  Circles
  represent a Mg\,II absorption threshold of $W_0=0.1$ \AA; squares
  represent $W_0=0.3$ \AA; and triangles represent $W_0=0.5$ \AA.  The
  data points are located at the median absolute $B$-band magnitudes
  of the galaxies with horrizontal errorbars representing the
  luminosity intervals of the subsamples.  Vertical errorbars
  represent the 68\% confidence interval.  For comparison, we have
  also included the limit of Gauthier \etal\ (2010) for Mg\,II
  absorbers with $\ewr\ge 0.5$ \AA\ around LRGs at $z\approx 0.5$.}
\end{center} 
\end{figure}

%  Considering the
%entire galaxy and absorber pair sample, we find that
%$\langle\kappa_{0.3}\rangle=0.59\pm 0.06$ for absorbers of
%$W_r(2796)>0.3$ \AA\ and $\langle\kappa_{0.1}\rangle=0.75\pm 0.05$ for
%absorbers of $W_r(2796)>0.1$ \AA\ at $\rho < 95\ h^{-1}$ kpc.  We note
%that including the outliers does not alter the results significantly.

\subsection{Density Profile of Extended Mg\,II Absorbing Gas}

The best-fit power-law model presented in Equation (11) and Figure 9
indicates that the observed absorber strength $W_r(2796)$ scales
inversely with increasing projected distance $\rho$ with a steep slope
of $\Delta\log\,W_r(2796)/\Delta\log\,\rho=-1.9$.  While the power-law
model appears to describe the observations well, a physical
interpretation of this steep anti-correlation is not straightforward.
Using a small sample of 13 galaxy--Mg\,II absorber pairs and 10
galaxies at $\rho<100\ h^{-1}$ kpc that do not give rise to Mg\,II
absorption to a sensitive upper limit ($W_r(2796)\apll 0.02$ \AA),
Chen \& Tinker (2008) showed that the anti-correlation between
$W_r(2796)$ and $\rho$, after accounting for $B$-band luminosity
distribution, is well described by either an isothermal density
profile or a Navarro-Frenk-White (NFW; Navarro \etal\ 1996) profile of
the absorbing clumps with a finite extent $R_{\rm gas}$.  An
isothermal density profile is motivated by the observed rotation
curves of nearby galaxies, while an NFW profile is found to represent
the density profiles of dark matter halos in high-resolution numerical
simulations.  Using the new sample of galaxy--Mg\,II absorber pairs
that is three times of the Chen \& Tinker (2008) sample, we examine
whether the anti-correlation displayed in Figure 9 is better described
by either an isothermal density profile or an NFW profile.

For an isothermal profile of gaseous clumps within a finite extent
$R_{\rm gas}$, the \ewr\ versus $\rho$ relation is characterized
following Chen \& Tinker (2008) as 
\begin{equation} 
\bar{W}_r^{\rm iso}(2796)
%\displaystyle\int_0^{\sqrt{R_{\rm gas}^2-\rho^2}}\frac{W_0}{(\rho^2+a_h^2)+l^2}\,dl \nonumber \\
=\frac{W_0}{\sqrt{\rho^2/a_h^2+1}}\tan^{-1}{\sqrt{\frac{R_{\rm gas}^2-\rho^2}{\rho^2+a_h^2}}}
\end{equation}
at $\rho\le R_{\rm gas}$ and $\ewr=0$ otherwise.  The core radius
$a_h$ is defined to be $a_h=0.2\,R_{\rm gas}$ and does not affect the
expected $\ewr$ at large $\rho$.  The extent of Mg\,II absorbing gas
scales with the luminosity of the absorbing galaxy according to
\begin{equation}
\frac{R_{\rm gas}}{R_{{\rm gas}*}}=\left(\frac{L_B}{L_{B_*}}\right)^{\beta}.
\end{equation}
Following the expectations of an isothermal model in Equations (16)
and (17), we perform the likelihood analysis described in Equations
(7) through (9) to find the best-fit values of $W_0$, $R_{\rm gas_*}$,
and $\beta$.  Excluding the outliers, the results of the likelihood
analysis show that the observations are best described by
$\log\,W_0^{\rm iso}=1.24\pm 0.03$,
\begin{equation}
\beta^{\rm iso}=0.35_{-0.04}^{+0.01},
\end{equation}
and
\begin{equation}
\log\,R_{\rm gas_*}^{\rm iso}=1.87\pm 0.01.
\end{equation}
The errors indicate the 95\% confidence intervals.  The best-fit
isothermal profile is also characterized by an intrinsic scatter of
$\sigma_c=0.104$ and an r.m.s.\ residual between the observed and
model Mg\,II absorber strengths of ${\rm
  r.m.s.}(\log\,W_r-\log\,\bar{W})=0.233$.

For an NFW profile, the absorber strength is expected to vary with
$\rho$ according to
\begin{equation}
\bar{W}_r^{\rm NFW}(2796)=\displaystyle\int_0^{\sqrt{R_{\rm gas}^2-\rho^2}}\frac{W_0\,r_s^3}{(r_s+\sqrt{l^2+\rho^2})^2\sqrt{l^2+\rho^2}}\,dl
\end{equation}
where $r_s$ is the scale radius.  We adopt $r_s\equiv R_{200}/15$ that
gives a halo concentration index of 15 (e.g.\ Dolag \etal\ 2004).  We
perform the likelihood analysis described in Equations (7) through (9)
and find $\log\,W_0^{\rm NFW}=-0.56\pm 0.05$,
\begin{equation}
\beta^{\rm NFW}=0.35\pm 0.03,
\end{equation}
and
\begin{equation}
\log\,R_{\rm gas_*}^{\rm NFW}=1.89_{-0.12}^{+0.05}.
\end{equation}
The errors indicate the 95\% confidence intervals.  The best-fit NFW
profile has associated intrinsic scatter of $\sigma_c=0.175$ and
r.m.s.\ residual between the observed and model Mg\,II absorber
strengths of ${\rm r.m.s.}(\log\,W_r-\log\,\bar{W})=0.253$.  

Figure 12 displays the best-fit models in comparison to observations.
While the isothermal model provides a somewhat better characterization
of the observations in terms of a minimum derived intrinsic scatter
and r.m.s.\ residual, both models result in a consistent best-fit
scaling relation of
\begin{equation}
\frac{R_{\rm gas}}{R_{\rm gas_*}}= \left(\frac{L_B}{L_{B_*}}\right)^{0.35},
\end{equation}
where the characteristic gaseous radius of an $L_*$ galaxy is found to be
\begin{equation}
R_{\rm gas_*}\approx 75\ h^{-1}\,{\rm kpc} 
\end{equation}
at $z\sim 0.25$.  

\begin{figure}
\begin{center}
\includegraphics[scale=0.4]{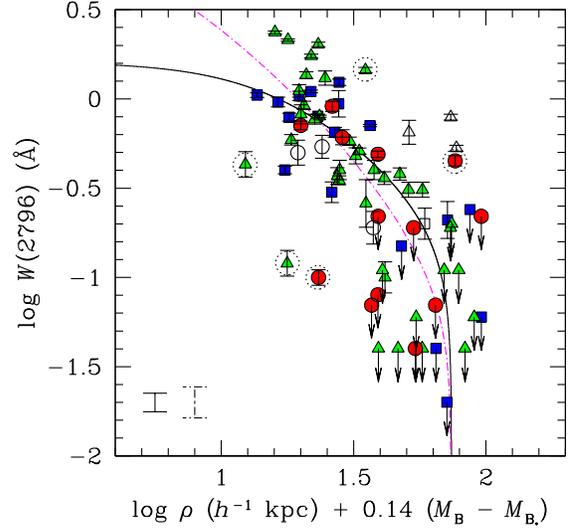}
\caption{Comparison of $W_r(2796)$ versus impact parameter accounting
  for scaling by $B$-band luminosity and redshift.  Symbols are the
  same as in Figure 8. The solid curve is the best-fit isothermal
  model and the dash-dotted curve is the best-fit NFW model, excluding
  five outliers (according to a 3-$\sigma$ clipping criterion; see \S\
  5.2) marked in dotted circles.  The errorbars in the lower-left
  corner indicate the intrinsic scatters of the adopted model profiles
  estimated based on the likelihood analysis.  We note that the
  residuals between observations and the best-fit model profile do not
  correlate with galaxy $B$-band luminosity, indicating that
  $W_0/R_{\rm gas}$ in Equation (16) does not depend on galaxy
  luminosity.}
\end{center} 
\end{figure}

It is clear that the scatter between the observations and the best-fit
isothermal model after accounting for the scaling relation with galaxy
luminosity remains large in Figure 12.  A natural expectation of the
isothermal model in Equation (16) is that $W_0$ depends on galaxy
properties, particularly on $L_B$ or mass, because more luminous (and
therefore more massive) galaxies possess more extended gaseous halos
and would exhibit stronger $W_0$ (e.g.\ Tinker \& Chen 2010).
However, we find no correlation between the residuals of
$(\log\,W_r-\log\,\bar{W})$ in Figure 12 and absolute $B$-band
magnitude of the galaxies, indicating that $W_0$ does not depend
strongly on galaxy luminosity/mass.  The lack of correlation suggests
that the mean absorber strength per halo does not vary with halo mass
and that the cool gas fraction declines with increasing halo mass in
the mass regime probed by our sample.

The best-fit scaling power in Equation (23) confirms the earlier
result of Chen \& Tinker (2008).  The best-fit characteristic gaseous
radius shown in Equation (24) is smaller than the earlier estimate of
$91_{-8}^{+3}\ h^{-1}$ kpc from a smaller galaxy sample, but they are
consistent to within 2-$\sigma$ error uncertainties.  The new galaxy
sample presented in this paper, which is three times larger than the
initial sample employed by Chen \& Tinker, has revealed a substantial
scatter, particularly at large radii.  The large uncertainty in
$R_{\rm gas_*}$ can be understood as due to the inherent degeneracy
between the spatial extent $R_{\rm gas}$ and diminishing covering
fraction $\kappa(\rho)$ of Mg\,II absorbing gas at large radii.  A
larger $R_{\rm gas_*}$ can be accommodated with a smaller gas covering
fraction.  For example, within $R_{\rm gas_*}=75\ h^{-1}$ kpc, 39 of
the 59 galaxies have an associated Mg\,II absorber of $\ewr\ge 0.3$
\AA.  This yields a gas covering fraction of $\kappa_{0.3}\approx
66$\% at $\rho' < 75\ h^{-1}$ kpc.  Within $R_{\rm gas_*}=91\ h^{-1}$
kpc, we find $\kappa_{0.3}\approx 61$\% at $\rho' < 91\ h^{-1}$ kpc
after excluding outliers.

In Chen \& Tinker (2008), the best-fit $R_{\rm gas_*}=91\ h^{-1}$ kpc
was close to the maximum impact separation of $r_0=100\ h^{-1}$ kpc in
the galaxy--QSO pair sample.  The selection criterion was imposed by
these authors to minimize the incidence of correlated galaxies through
large-scale clustering (see their Section 2 for discussion).  Chen \&
Tinker examined the possible bias in the best-fit $R_{{\rm gas}_*}$ due
to the imposed $r_0$ selection, and found that the best-fit $R_{\rm
  gas_*}$ was insensitive to the choice of $r_0$.  Namely, they
obtained a consistent estimate of $R_{\rm gas_*}$ for a smaller $r_0$.
However, choosing a smaller $r_0$ reduced the already small sample
further and the parameters became poorly constrained.

In the present analysis, we have included galaxy--QSO pairs with
impact separation as large as $\rho\approx 120\ h^{-1}$ kpc (Figure
3).  The best-fit gaseous radius of $R_{\rm gas_*}=75\ h^{-1}$ kpc is
significantly below the the maximum separation in the pair sample,
although the majority of the data (galaxies with intrinsic colors
consistent with present-day disks and irregular's/starbursts) lie at
$\rho<75\ h^{-1}$ kpc.  To examine whether the best-fit $R_{\rm
  gas_*}$ is sensitive to the range of impact parameter covered by the
pair sample, we repeat the likelihood analysis adopting only pairs
with $\rho<75\ h^{-1}$ kpc.  We find that the best-fit results remain
the same.

In summary, we find based on a sample of 71 'isolated' galaxies that
the extent of cool gas as probed by the Mg\,II absorption features
scales with galaxy $B$-band luminosity according to $R_{\rm
gas}\propto\,L_B^{0.35}$, consistent with earlier results obtained
using smaller samples.  As discussed in Chen \& Tinker (2008), the
scaling relation, when combined with a fiducial mass-to-light ratio
determined from galaxy halo occupation studies of wide-field galaxy
survey data, indicates that the gaseous radius is a constant fraction
of the halo radius over the mass scale of
$M_h=10^{10.6}-10^{13}\,\hmsol$ sampled in our
observations\footnote{The halo occupation analysis of 2dFGRS galaxies
from Tinker \etal\ (2007) showed that the relationship between $M_h$
and luminosity for galaxies of $L_{b_J}\lesssim L_{*}$ at $z\sim 0.1$
is $M_h = 10^{12.5}\,(L_B/L_{B_*})^{1.3}\,\hmsol$.  This monotonic
relationship is appropriate for galaxies that reside at the {\it
centers} of their dark matter halos and are the brightest galaxy in
the halo.  A similar scaling relation was obtained, $M_h =
10^{11.9}\,(L_B/L_{B_*})^{1.3}\,\hmsol$, for DEEP2 galaxies at $z\sim
1$ by Zheng \etal\ (2007).  Interpolating in $\log\,(1+z)$, we find
that galaxies of $M_{B_*}-5\log\,h=-19.8$ on average reside in halos
of $10^{12.4}\,\hmsol$ at $z\sim 0.25$.}.  In addition, we find that
the residuals of observed and best-fit $W_r(2796)$ do not correlate
with galaxy absolute $B$-band magnitude, implying a declined cold gas
fraction with increasing halos mass in the mass regime probed by our
sample.

\subsection{The Origin of Extended Cool Gas and Implications for the Baryon Content of Galactic Halos}

The presence of a finite extent $R_{\rm gas}$ of Mg\,II absorbing gas
is similar to what is found for extended C\,IV (Chen \etal\ 2001) and
also for OVI absorbing gas recently published by Chen \& Mulchaey
(2009) around galaxies.  The origin of such a finite boundary for
metal-line absorbers can be interpreted as due to a halo fountain
phenomenon (c.f.\ Bregman 1980), according to which the gaseous radius
is driven by the finite distance the outflowing material can travel
from an early episode of starburst.  We find this scenario unlikely
because of the broad range of intrinsic colors covered by our galaxy
sample, but a detailed stellar population synthesis to investigate the
star formation history will shed more light on this issue.

Alternatively, the finite boundary of absorbing clouds probed by these
metal-line absorbers can be understood as a critical radius below
which cool clouds can form and survive in an otherwise hot medium.
This two-phase model to interpret QSO absorption line systems was
formulated in Mo \& Miralda-Escud\'e (1996) and later re-visited by
Maller \& Bullock (2004).  This is also shown in recent
high-resolution numerical simulations of Milky Way type halos, in
which accreted cold streams are disrupted by shocks but the remaining
gas overdensities serve as the 'seeds' necessary to form cool gaseous
clouds through thermal instabilities within $\approx 1/3$ of the halo
radius (e.g.\ Kere{\v s} \& Hernquist 2009).  

The formation and presence of such cool clouds in a hot halo have
several important implications for the studies of galaxy formation and
evolution.  For example, the confining hot medium around the cool
clouds is expected to have a low density and long cooling time,
reducing the ``overcooling problem'' in standard galaxy formation
models (e.g.\ Maller \& Bullock 2004; Kaufmann \etal\ 2009).  In
addition, these clouds may provide the fuels necessary to support
continuous star formation near the center of galactic halos (e.g.\
Binney \etal\ 2000).  Furthermore, the cool clouds together with the
confining hot medium offer a reservoir for missing baryons in
individual galactic halos (e.g.\ McGaugh \etal\ 2010).  A nominal
candidate for these cool clouds in our Milky Way Halo is the high
velocity clouds (HVCs) of neutral hydrogen column density $N({\rm
H\,I})>10^{18}$ \cmjj\ seen in all-sky 21~cm observations.  However,
unknown distances of these HVCs prohibit an accurate measurement of
their total gas mass (e.g.\ Putman 2006).

High-resolution spectra of strong Mg\,II absorbers have revealed the
multi-component nature of the absorbing gas, with \ewr\ roughly
proportional to the number of components in the system (e.g.\
Petitjean \& Bergeron 1990; Prochter et al. 2006).  These discrete
Mg\,II absorbing components are natural candidates of the condensed
cool clouds within $\sim 75\ h^{-1}$ kpc radius of known galaxies.
Interpreting the intrinsic scatter $\sigma_c$ as due to Poisson noise
in the number of cool clumps intercepted along a line of sight allows
us to estimate the number of absorption clumps per galactic halo per
sightline, $n^{\rm clump}$.  In this simple model, each absorber is
characterized as $W_r=n^{\rm clump}\times k$, where $k$ is the mean
absorption equivalent width per absorbing component.  Following
Poisson counting statistics, the intrinsic scatter $\sigma_{c}\equiv
\sigma_{c,W_r}/(W_r\,\ln\,10)$ is related to $n^{\rm clump}$ according
to
\begin{equation}
\sigma_{c} = \frac{1}{\ln\,10}\frac{\sqrt{n^{\rm clump}+1}}{n^{\rm clump}}
\end{equation}
For the isothermal model, we have determined a best-fit intrinsic
scatter of $\sigma_c=0.104$, which leads to a mean number of absorbing
clumps of
\begin{equation}
  n^{\rm clump} \sim 18
\end{equation}
and a mean absorption equivalent width per clump of $k = 0.04$ \AA\ at
a median projected distance of $\langle\rho\rangle=26\ h^{-1}$ kpc.
The corresponding Mg\,II absorbing column density per clump is $N({\rm
Mg\,II})\approx 10^{12}$ \cmjj.  We note that if other factors
contribute to the observed $\sigma_c$, then the corresponding Poisson
noise is smaller and the inferred $n^{\rm clump}$ will be bigger.

To estimate the total baryonic mass contained in these cool clouds
traced by Mg\,II absorbers, we first assume a mean clump size of 1 kpc
(consistent with the size of H\,I clouds seen around M31; Westmeier
\etal\ 2008), a mean metallicity of 1/10 solar (consistent with what
is seen in the HVC Complex C; see Thom \etal\ 2008 for a list of
references), and a mean ionization fraction of $f_{\rm Mg^+}=0.1$ for
photo-ionized clouds (see Figure 6 of Chen \& Tinker 2008).  We derive
a mean gas density of $n_{\rm H}\approx 10^{-3}\ {\rm cm}^{-3}$.
Assuming a spherical shape for the clumps leads to a mean clump mass
of $M_b^{\rm clump}=1.7\times 10^4\,{\rm M}_\odot$.  Next, we adopt an
isothermal density distribution of the clumps (Equation 16) and derive
a total number of clumps $N^{\rm clump}=2\times 10^5\, h^{-2}$ within
$R_{\rm gas}=75\ h^{-1}$ kpc.  The total baryonic mass in these cool
clumps is found to be $M_b\sim 3\times 10^9\,h^{-2}\,{\rm M}_\odot$,
comparable to the total cold gas content seen in the Milky Way disk
(e.g.\ Flynn \etal\ 2006).  The inferred total mass would be still
larger if $n^{\rm clump}$ increases as the Poisson noise decreases.

If we further assume that it takes a free-fall time for the clumps to
reach the center of the halo, we infer a cool gas accretion rate of
$\dot{M}_{\rm gas}\sim 3\,{\rm M}_\odot\,{\rm yr}^{-1}$.  The inferred
gas accretion rate is interestingly close to the star formation rate
seen in the Milky Way disk (e.g.\ Robitaille \& Whitney 2010).
However, we note two issues in this simple picture.

The first issue concerns the lifetime of these cool clumps relative to
the time it takes for them to reach the inner few kpc of the host
halo.  Previous studies have shown that clouds of mass $M_c\apll
10^5\,{\rm M}_\odot$ cannot survive thermal conduction and ram
pressure as they move through a low-density, hot medium (see e.g.\ Mo
\& Miralda-Escud\'e 1996; Maller \& Bullock 2004; Heitsch \& Putman
2009).  To alleviate this problem, the clumps will need to have lower
metallicities and/or bigger sizes than what we have assumed.  For
example, if we assume a mean clump size of 2 kpc and a mean
metallicity of 0.05 solar, then we derive a mean clump mass of
$M_b^{\rm clump}=1.3\times 10^5\,{\rm M}_\odot$.  The total number of
clumps would be $N^{\rm clump}=5\times 10^4\, h^{-2}$ and the total
cool gas mass would be $M_b\sim 6.5\times 10^9\,h^{-2}\,{\rm M}_\odot$
at $r\le 75\ h^{-1}$ kpc.

Second, we note that the expected H\,I column density of these Mg\,II
absorbing clumps is $N({\rm H\,I})\sim 10^{16}$ \cmjj, below the
$N({\rm H\,I})$ threshold provided by 21~cm observations.  The H\,I
content is therefore too low for these clumps to be the distant
counterparts of the local HVCs.  Nevertheless, the lack of strong
($\ewr\ge 0.5$ \AA) Mg\,II absorbers beyond $30\ h^{-1}$ kpc (Figure
10) which have have associated H\,I of $N({\rm HI})> 10^{18}$ \cmjj\
(Rao \etal\ 2006), constrains the distances of the Milky Way HVCs at
$d\apll 50$ kpc.  Within this close distance range, the implied total
gas mass in the HVCs is expected to be $M_b^{\rm HVC}<10^9\,{\rm
M}_\odot$ (Putman 2006).

The exercise presented above shows that the cool clouds that give rise
to Mg\,II absorbers and the implied presence of a hot confining medium
can provide a substantial reservoir for missing baryons in individual
galactic halos (see also Kaufmann \etal\ 2009).  The presence of
extended Mg\,II absorbing halos around galaxies of a broad range of
intrinsic colors argues against an outflow origin of these absorbers.
Under an infall scenario, we show that these cool clouds provide
sufficient fuels to support continuous star formation in their host
galaxies.

\subsection{Comparison with Other Studies}

The large covering fraction of Mg\,II absorbing gas (\S\ 6.1) and the
scaling relation between gaseous extent and galaxy $B$-band luminosity
(\S\ 6.2) represent the empirical evidence supporting the halo
occupation model developed in Tinker \& Chen (2008, 2009; see also
Gauthier \etal\ 2009).  Here we compare the results of our studies
with recent surveys by other groups.

First, we have noted the discrepancy between the high gas covering
fraction around $\sim L_*$ galaxies and the low incidence of Mg\,II
absorbers around luminous red galaxies studied in Gauthier \etal\
(2010).  We have interpreted the contrast as due to the different halo
mass scales probed by the two galaxy samples.  Cosmological
simulations have shown that the growth of galaxies progresses
primarily through stable accretion of intergalactic gas (e.g.\
Kere\v{s} \etal\ 2005; 2009).  In low-mass halos ($M_h < M_h^{\rm
  crit}=10^{11.5}\,\hmsol$), accretion proceeds to the center of the
halo through dense cold streams (Dekel \& Birnboim 2006; Kere\v{s}
\etal\ 2005; 2009).  As galaxies grow larger in mass ($M_h > M_h^{\rm
  crit}=10^{11.5}\,\hmsol$), both numerical simulations and analytic
models show that a progressively larger fraction of the accreted gas
is shock heated to the virial temperature.  Some fraction of the
shock-heated gas may be able to cool in the inner region where gas
density is high.  The majority of the galaxies presented in this paper
are on the transitional mass scale, $M_h = 10^{11-12.5}\,\hmsol$.  In
contrast, the luminous red galaxies studied by Gauthier \etal\ (2010)
are in a higher mass regime of $M_h>10^{13}\,\hmsol$, where a
diminishing amount of cold gas is expected because the cooling time is
long.  The differential gas covering fraction found in our galaxies
and in those luminous red galaxies is therefore consistent with
theoretical expectations.

Kacprzak \etal\ (2008) have also studied the extent and covering
fraction of Mg\,II absorbing gas based on a sample of 37 galaxies at
$z=0.3-1$ that are found in the vicinity of a known Mg\,II absorber.
While they did not see a clear correlation between the extent of
Mg\,II absorbing gas and galaxy luminosity, they found that for a
scaling power of $\beta=0.18-0.58$ and a characteristic gaseous extent
of $R_*=56-105\ h^{-1}$ kpc the covering fraction of Mg\,II absorbers
with $\ewr\ge 0.3$ \AA\ is no more than 70\%.  Our best-fit isothermal
model indicates that the characteristic gaseous radius at $\ewr=0.3$
\AA\ is $R_{\rm gas_*}(W_r=0.3\AA)=47\ h^{-1}$ kpc (Figure 11), less
than the gaseous radius considered by Kacprzak \etal.  Excluding the
outliers, we find a mean covering fraction of $\kappa_{0.3}[\rho'<47\
h^{-1}\,{\rm kpc}]\approx 76$\% for Mg\,II absorbers with $\ewr\ge
0.3$ \AA.  Including the outliers, we find a mean covering fraction of
$\kappa_{0.3}[\rho'<47\ h^{-1}\,{\rm kpc}]\approx 78$\% for Mg\,II
absorbers with $\ewr\ge 0.3$ \AA.  The gas covering fraction found in
our sample is somewhat higher than the parameter space explored by
Kacprzak \etal, but this can be explained by the larger gaseous radius
considered by these authors.

In addition, Barton \& Cooke (2009) have recently carried out a search
of coincident Mg\,II absorbers in the vicinity of 20 luminous galaxies
($M_r-5\,\log\,h\le -20.5$) identified at $z\sim 0.1$ in the SDSS
galaxy spectroscopic sample.  The authors reported a low gas covering
fraction of $\kappa_{0.3}[\rho<75\ h^{-1}\ {\rm kpc}]\apll 0.4$ and
$\kappa_{0.3}[\rho<35\ h^{-1}\ {\rm kpc}]\sim 0.25$ for Mg\,II
absorbers of $\ewr\ge 0.3$ \AA.  These results are inconsistent with
the high gas covering fraction found in our survey.  The design of the
Barton \& Cooke survey is very similar to ours presented here, with
only small differences in the targeted redshift range ($z\sim 0.1$)
and luminosity scale ($>L_*$).  The discrepancy in the observed gas
covering fraction therefore requires further perusal.

A fundamental difference in our respective measurements is that Barton
\& Cooke excluded observed strong Mg\,II absorbers in the covering
fraction calculation if the QSO spectrum did not have sufficient $S/N$
for identifying absorbers of $\ewr=0.3$ \AA.  Of the six galaxies at
$\rho \le 35\ h^{-1}$ kpc, three have associated Mg\,II absorbers of
$\ewr=1.93\pm 0.18$ \AA, $\ewr=1.91\pm 0.10$ \AA, and $\ewr=0.71\pm
0.15$ \AA, respectively, and three do not have corresonding Mg\,II
absorbers to 3-$\sigma$ limits of $\ewr<0.3$ \AA.  The authors
excluded from the covering fraction calculation two strong absorbers,
$\ewr=1.93\pm 0.18$ \AA\ and $\ewr=0.71\pm 0.15$ \AA, and derived
$\kappa_{0.3}[\rho<35\ h^{-1}\ {\rm kpc}]\sim 0.25$ based on the
remaining four galaxies.

We note that the goal of our study is to determine the incidence of
Mg\,II absorbers at the location of a known galaxy.  Absorbers that
are observed to have $\ewr=0.71\pm0.15$ \AA\ and $\ewr=1.93\pm 0.18$
\AA\ confirm the presence of such absorbing gas in large quantities,
despite the fact that the QSO spectrum may not offer the sensitivities
required for uncovering weaker absorbers of $\ewr\sim 0.3$ \AA\ at
high confidence levels.  Excluding these sightlines imposes a bias
that would result in an underestimate of the gas covering fraction.

We have retrieved from the SDSS data archive the galaxy photometric
and spectroscopic data in Barton \& Cooke (2009), and determined the
rest-frame $B$-band absolute magnitude and $B_{AB}-R_{AB}$ color for
each of the 20 galaxies according to the procedures described in \S\
4.1.  The rest-frame absolute $B$-band magnitudes of these galaxies
span a range from $M_{B}-5\log\,h=-19.6$ to $M_{B}-5\log\,h=-20.7$
with a median of $\langle M_{B}-5\log\,h\rangle=-20.3$.  The
rest-frame $B_{AB}-R_{AB}$ colors range from $B_{AB}-R_{AB}\approx
0.6$ to $B_{AB}-R_{AB}\approx 1.2$ with a median of $\langle
B_{AB}-R_{AB}\rangle_{\rm med}\approx 1.0$.  Figure 13 displays the
observed \ewr\ versus $\rho$ relation for the Barton \& Cooke sample,
superimposed on top of our own data.  We have converted the 3-$\sigma$
upper limits published by these authors to 2-$\sigma$ upper limits in
order to be consistent with our own measurements.

\begin{figure}
\begin{center}
\includegraphics[scale=0.4]{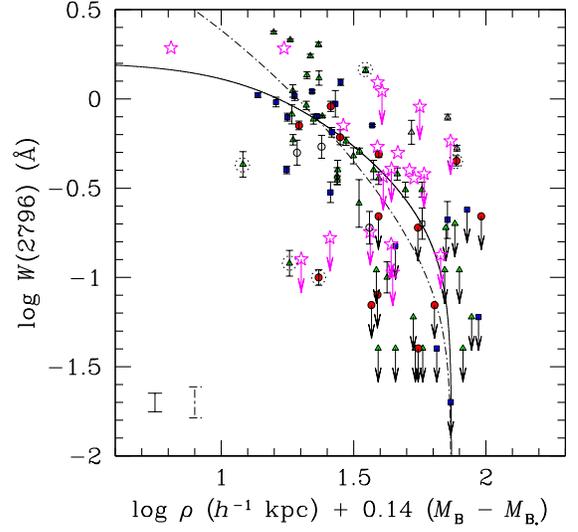}
\caption{Comparison of the observed $\ewr$ versus $\rho$ correlation
  for galaxies in our sample and those published in Barton \& Cooke
  (2009).  We have applied the scaling relation of Equation (23) for
  both samples.  Symbols and best-fit models for our sample are the
  same as in Figure 11.  The galaxies of Barton \& Cooke are shown in
  star symbols.  We note that the galaxy at
  $\log\,\rho'=\log\,\rho+0.14\,(M_B-M_{B_*})\approx 1.6$ with a
  strong Mg\,II absorber of $\ewr=1.24$ (SDSSJ144033.82$+$044830.9) in
  the Barton \& Cooke sample has been found to have a companion galaxy
  at $\rho=18.6\ h^{-1}$ kpc.  The absorber would therefore be
  considered as in a 'group' environment.}
\end{center} 
\end{figure}

Figure 13 shows that with the exception of two outliers at
$\rho'\approx 25\ h^{-1}$ kpc, the Barton \& Cooke sample is in
general agreement with what is seen in our sample.  Six of the 12
upper limits in the Barton \& Cooke sample do not have sufficient
sensitivities for constraining the presence/absence of Mg\,II
absorbing gas at the level of $\ewr\sim 0.3$ \AA.  In addition, the
galaxy at $\log\,\rho'=\log\,\rho+0.14\,(M_B-M_{B_*})\approx 1.6$ with
a strong Mg\,II absorber of $\ewr=1.24$ (SDSSJ144033.82$+$044830.9)
was later found to have a companion galaxy at $\rho=18.6\ h^{-1}$ kpc.
The absorber would therefore be considered as in a 'group'
environment.  The galaxy at $\rho'\approx 20\ h^{-1}$ kpc with a
2-$\sigma$ upper limit of $\ewr<0.13$ \AA\ (SDSSJ151541.23$+$334739.4)
has a luminous companion (SDSSJ151536.18$+$334743.6) at projected
distance $< 100 \ h^{-1}$ kpc and velocity separation
$\Delta\,v\approx 27$ \kms\ away.  This pair also respresents a
'group' that we have treated separately in our analysis.

Considering only pairs between 'isolated' galaxies and QSOs for which
sensitive limits for the corresponding Mg\,II absorption are available
leads to a sample of 12 galaxies with sensitive constraints for the
corresponding Mg\,II absorption features.  We find three of the four
galaxies at $\rho'<35\ h^{-1}$ kpc have associated Mg\,II absorbers of
$\ewr>0.3$ \AA, yielding $\kappa_{0.3}[\rho'<35\ h^{-1}\ {\rm
kpc}]\approx 0.75$.  We find seven of the 12 galaxies at $\rho'<75\
h^{-1}$ kpc have associated Mg\,II absorbers of $\ewr>0.3$ \AA,
yielding $\kappa_{0.3}[\rho'<75\ h^{-1}\ {\rm kpc}]\approx 0.58$.
These values are consistent with our estimated gas covering fraction
to within 1-$\sigma$ uncertainties.  We therefore conclude that the
observations presented in Barton \& Cooke (2009) are consistent with
the high gas covering fraction seen in our sample.

Finally, Zibetti \etal\ (2007) studied the nature of Mg\,II absorbing
galaxies at $z=0.37-1$ based on stacked images of QSOs with known
foreground Mg\,II absorbers of $\ewr > 0.8$ \AA.  After subtracting
the QSO point spread function in the stacked images, these authors
obtained an angular averaged image of Mg\,II absorbing galaxies out to
100 kpc.  They found that the luminosity-weighted mean colors of the
extended emission is consistent with present-day intermediate-type
galaxies.  In addition, the authors found that weak absorbers of
$\ewr<1.1$ \AA\ originate primarily from red passive galaxies while
stronger absorbers display bluer colors consistent with star-forming
galaxies.  Based on the differential color distribution, Zibetti
\etal\ argued that the origin of strong Mg\,II absorbers might be
better explained by models of metal-enriched gas outflows from
star-forming/bursting galaxies.

Our analysis in \S\ 5.3 shows that the observed Mg\,II absorber
strength does not depend strongly on galaxy intrinsic color, which
appears to contradict the observations of Zibetti \etal.  To
understand the discrepant results, we first recall based on the
observed $\ewr$ versus $\rho$ anti-correlation in Figure 8 that the
mean impact parameter of the absorbing galaxies for weak absorbers is
expected to be larger than that for stronger absorbers.  In addition,
Figure 8 also shows that for a sample of randomly selected QSO--galaxy
pairs in our study, the fraction of red galaxies is found to be a
factor of two larger at larger impact parameters $\rho>30\ h^{-1}$ kpc
than at closer distances.  This is not due to selection biases,
because we did not preferentially select blue galaxies at smaller
impact parameters.  Instead, this can be understood by considering the
fact that redder galaxies are on average more luminous and have a
lower space density than bluer and on average fainter galaxies (e.g.\
Faber \etal\ 2007).  The increased incidence of red galaxies at larger
impact parameter, combined with the luminosity-weighted nature of the
image stacking procedure, is expected to skew the stacked images of
weaker absorbers toward redder colors.  Whether or not this is
sufficient to explain the observed color gradient between strong and
weak Mg\,II absorbers in the Zibetti \etal\ sample requires a detailed
simulation that takes into account such intrinsic weighting by the
space density and luminosity of the galaxies.  Because such analysis
has not been performed, it is not clear whether the respective results
are inconsistent with each other.  Nevertheless, the conclusion in
favor of strong Mg\,II absorbers being better explained by starburst
outflows based on the observed color gradient should be viewed with
caution.

\section{SUMMARY}

We are conducting a survey of Mg\,II absorbers in the spectra of
background QSOs that are within close angular distances to a
foreground galaxy at $z<0.5$, using the Magellan Echellette
Spectrograph.  The goal of this project is to establish a
statistically significant sample ($N\apg 500$) of galaxies with
sensitive constraints on their surrounding gaseous halos for measuring
the clustering amplitude of Mg\,II absorbers at $\langle z\rangle=0.2$
and for a comprehensive study of how the properties of extended cool
gas correlate with known stellar and ISM properties of the host
galaxies.

Here we report the results of the first-year survey data.  We have so
far established a spectroscopic sample of 94 galaxies at a median
redshift of $\langle z\rangle = 0.2357$ in fields around 70 distant
background QSOs ($z_{\rm QSO}>0.6$).  The galaxies span a broad range
in the intrinsic $B_{AB}-R_{AB}$ color, from $B_{AB}-R_{AB}\approx 0$
to $B_{AB}-R_{AB}\approx 1.5$, and a broad range in the rest-frame
absolute $B$-band magnitude, from $M_{B}-5\log\,h=-16.22$ to
$M_{B}-5\log\,h=-22.5$.  Of the 94 galaxies, 17 are found with at
least one neighbor at close projected distances $\rho<{\hat R}_{\rm
  gas}$ (defined in Equation 1) and velocity separations
$\Delta\,v<300$ \kms; one occurs at $z=0.0934$, outside of the
redshift range allow by the spectral coverage of MagE for the Mg\,II
absorber search; and five do not have sensitive limits for $W_r(2796)$
due to contamination by other absorption features.  Excluding these
galaxies leads to a sample of 71 'isolated' galaxies at $\rho\apll
120\ h^{-1}$ kpc from the line of sight of a background QSO.  Of the
71 galaxies, we find that 47 have corresponding Mg\,II absorbers in
the spectra of background QSOs with absorber strengths varying from
$W_r(2796)=0.1$ \AA\ to $W_r(2796)=2.34$ \AA, and 24 do not give rise
to Mg\,II absorption to a sensitive upper limit.

We examine the correlation between the properties of galaxies and
Mg\,II absorption strengths, using the sample of 71 'isolated'
galaxies.  The main results of our study are summarized below.

1. Based on a likelihood analysis, we confirm previous results that
the Mg\,II absorber strength $W_r(2796)$ declines with increasing
galaxy impact parameter and that the $W_r(2796)$ vs.\ $\rho$
anti-correlation is further strengthened after accounting for the
scaling relation with galaxy $B$-band luminosity.  The significantly
improved anti-correlation indicates that more luminous galaxies are
surrounded by more extended Mg\,II halos.

2.  While galaxies from a 'group' environment appear to occupy a
similar $W_r(2796)$ versus $\rho$ parameter space with 'isolated'
galaxies, they do not exhibit a strong inverse correlation.

3. Including intrinsic $B_{AB}-R_{AB}$ color does not improve the
observed $W_r(2796)$ vs.\ $\rho$ anti-correlation.  The broad range of
intrinsic colors in our sample, indicating a broad range of galaxy
types from E/SO to starburst, shows that extended Mg\,II halos are a
common and generic feature of ordinary galaxies.  It suggests a lack
of physical connection between the origin of extended Mg\,II halos and
recent star formation history of the galaxies.

4. Including galaxy redshift does not improve the $W_r(2796)$ vs.\
$\rho$ anti-correlation.  Therefore, our galaxy sample reveals little
evolution in the extended Mg\,II halos between $z\approx 0.5$ and
$z\approx 0.1$.

5. The extended halos of Mg\,II absorbing gas around ordinary galaxies
at $z\sim 0.25$ are best described by $R_{\rm gas}/R_{\rm gas_*}=
(L_B/L_{B_*})^{0.35\pm 0.03}$ and $R_{\rm gas_*}\approx 75\ h^{-1}$
kpc, indicating that gaseous radius is a constant fraction of halo
radius over the mass scale of $M_h=10^{10.6}-10^{13}\,\hmsol$ sampled
in our observations.  No correlation is found between the number
density of cool gas clouds and galaxy luminosity.

6. The covering fraction of extended Mg\,II absorbing gas is high.
Within $R_{\rm gas}$, we find a mean covering fraction of
$\langle\kappa_{0.3}\rangle\approx 70$\% for absorbers of $\ewr\ge
0.3$ \AA\ and $\langle\kappa_{0.1}\rangle\approx 80$\% for absorbers
of $\ewr\ge 0.1$ \AA.

7. Interpreting the remaining scatter in the luminosity-scaled \ewr\
versus $\rho$ anti-correlation as due to Poisson noise in the number
of intercepted absorbing clumps along the sightline, we estimate that
a sightline at $\langle\rho\rangle=26\ h^{-1}$ kpc intercepts on
average a total of $n^{\rm clump}\sim 18$ clumps through the gaseous
halo.  Adopting an isothermal density distribution of the clumps, we
further estimate a total baryonic mass of the cool clumps probed by
Mg\,II features of $M_b\sim (3-6)\times 10^9\,h^{-2}\,{\rm M}_\odot$.
The inferred total mass would be still larger if other factors
contribute to the observed scatter in the mean $\ewr$ versus $\rho$
relation.

The results from our survey are consistent with previous studies based
on smaller samples.  The large covering fraction of Mg\,II absorbing
gas and the scaling relation between gaseous extent and galaxy
$B$-band luminosity provide the empirical support for the halo
occupation model developed in Tinker \& Chen (2008, 2009) and in
Gauthier \etal\ (2009).

\acknowledgments

We are grateful to the SDSS collaboration for producing and
maintaining the SDSS public data archive.  We thank Niraj Welikala for
providing photometric measurements of galaxy
SDSSJ090519.72$+$084914.02.  We also thank Betsy Barton, Nick Gnedin,
Andrey Kravtsov, and Mary Putman for helpful discussions.
H.-W.C. acknowledges partial support from NASA Long Term Space
Astrophysics grant NNG06GC36G and an NSF grant AST-0607510.
J.-R.G. acknowledges partial support from a Brinson Foundation
Predoctoral Fellowship.  Funding for the SDSS and SDSS-II was provided
by the Alfred P. Sloan Foundation, the Participating Institutions, the
National Science Foundation, the U.S. Department of Energy, the
National Aeronautics and Space Administration, the Japanese
Monbukagakusho, the Max Planck Society, and the Higher Education
Funding Council for England. The SDSS was managed by the Astrophysical
Research Consortium for the Participating Institutions.

%\bibliography{ovi-statistics}
%\bibliographystyle{apj}

%\clearpage

%\begin{center}
\begin{deluxetable}{p{1.5in}ccccrrc}
\tablewidth{0pc}
\tablecaption{Summary of Faint Galaxy Spectroscopy}
\tabletypesize{\tiny}
\tablehead{ \colhead{ID} & \colhead{RA(J2000)} & \colhead{Dec(J2000)} & \colhead{$z_{\rm phot}$} & \colhead{$r'$} & \colhead{Instrument} & \colhead{Exptime} & \colhead{UT Date}}
\startdata
SDSSJ003339.66$-$005518.36 & 00:33:39.66 & $-$00:55:18.36 & $0.18\pm0.06$ & 19.6 & DIS & $3\times 1200$ & 2008-11-23 \\
SDSSJ003339.85$-$005522.36 & 00:33:39.85 & $-$00:55:22.36 & $0.16\pm0.03$ & 18.8 & MagE & $2\times 300$ & 2008-09-23 \\
SDSSJ003341.47$-$005522.79 & 00:33:41.47 & $-$00:55:22.79 & $0.28\pm0.12$ & 20.5 & DIS & $2\times 1800+1200$ & 2008-10-24 \\
SDSSJ003407.78$-$085453.28 & 00:34:07.78 & $-$08:54:53.28 & $0.35\pm0.12$ & 21.3 & MagE & 600 & 2008-09-24 \\
SDSSJ003412.85$-$010019.79 & 00:34:12.85 & $-$01:00:19.79 & $0.25\pm0.05$ & 20.1 & MagE & $2\times 900$ & 2008-09-23 \\
SDSSJ003414.49$-$005927.49 & 00:34:14.49 & $-$00:59:27.49 & $0.12\pm0.01$ & 17.2 & SDSS & ... & ... \\
SDSSJ010136.52$-$005016.44 & 01:01:36.52 & $-$00:50:16.44 & $0.22\pm0.06$ & 19.4 & MagE & $2\times 600$ & 2008-09-24 \\
SDSSJ010155.80$-$084408.74 & 01:01:55.80 & $-$08:44:08.74 & $0.26\pm0.10$ & 20.1 & MagE & 600 & 2008-09-25 \\
SDSSJ010351.82$+$003740.77 & 01:03:51.82 & $+$00:37:40.77 & $0.29\pm0.13$ & 21.3 & MagE & $2\times 600$ & 2008-09-25 \\
SDSSJ021558.84$-$011131.23 & 02:15:58.84 & $-$01:11:31.23 & $0.16\pm0.07$ & 19.3 & DIS & $2\times 1800+1300$ & 2008-11-01 \\
SDSSJ022949.97$-$074255.88 & 02:29:49.97 & $-$07:42:55.88 & $0.27\pm0.17$ & 20.9 & DIS & $3\times 1800$ & 2008-11-30 \\
SDSSJ024127.75$-$004517.04 & 02:41:27.75 & $-$00:45:17.04 & $0.14\pm0.04$ & 18.6 & DIS & $1800+1500$ & 2008-11-17 \\
SDSSJ032230.27$+$003712.72 & 03:22:30.27 & $+$00:37:12.72 & $0.20\pm0.03$ & 17.5 & DIS & $2\times 900$ & 2008-11-23 \\
SDSSJ032232.55$+$003644.68 & 03:22:32.55 & $+$00:36:44.68 & $0.38\pm0.12$ & 21.2 & DIS & $2\times 1800$ & 2008-12-22 \\
SDSSJ035241.99$+$001317.13 & 03:52:41.99 & $+$00:13:17.13 & $0.28\pm0.08$ & 20.3 & MagE & $2\times 600$ & 2008-09-24 \\
SDSSJ040404.51$-$060709.46 & 04:04:04.51 & $-$06:07:09.46 & $0.32\pm0.10$ & 20.2 & MagE & 1200 & 2009-02-22 \\
SDSSJ074527.07$+$191959.90 & 07:45:27.07 & $+$19:19:59.90 & $0.36\pm0.03$ & 19.7 & DIS & $2\times 1800$ & 2009-02-18 \\
SDSSJ074527.22$+$192003.88 & 07:45:27.22 & $+$19:20:03.88 & $0.28\pm0.09$ & 20.2 & DIS & $2\times 1800$ & 2009-02-18 \\
SDSSJ075001.34$+$161301.92 & 07:50:01.34 & $+$16:13:01.92 & $0.17\pm0.11$ & 20.6 & MagE & $300+600$ & 2009-03-28 \\
SDSSJ075450.11$+$185005.28 & 07:54:50.11 & $+$18:50:05.28 & $0.33\pm0.04$ & 19.7 & DIS & $2\times 1200$ & 2009-02-18 \\
SDSSJ075525.13$+$172825.79 & 07:55:25.13 & $+$17:28:25.79 & $0.18\pm0.07$ & 19.3 & DIS & $2\times 1200$ & 2009-02-18 \\
SDSSJ080005.11$+$184933.31 & 08:00:05.11 & $+$18:49:33.31 & $0.28\pm0.08$ & 19.9 & DIS & 1800 & 2009-03-19 \\
SDSSJ082340.56$+$074751.07 & 08:23:40.56 & $+$07:47:51.07 & $0.18\pm0.04$ & 18.5 & DIS & 1800 & 2009-01-05 \\
SDSSJ083218.55$+$043337.81 & 08:32:18.55 & $+$04:33:37.81 & $0.17\pm0.04$ & 18.1 & DIS & 1200 & 2009-03-19 \\
SDSSJ083218.77$+$043346.58 & 08:32:18.77 & $+$04:33:46.58 & $0.17\pm0.02$ & 17.4 & DIS & 1200 & 2009-03-19 \\
SDSSJ083221.60$+$043359.74 & 08:32:21.60 & $+$04:33:59.74 & $0.16\pm0.08$ & 19.4 & DIS & $2\times 1200$ & 2009-02-18 \\
SDSSJ084120.59$+$012628.85 & 08:41:20.59 & $+$01:26:28.85 & $0.20\pm0.11$ & 20.0 & DIS & $2\times 600$ & 2009-02-18 \\
SDSSJ084455.58$+$004718.15 & 08:44:55.58 & $+$00:47:18.15 & $0.17\pm0.06$ & 18.8 & DIS & 1800 & 2009-01-05 \\
SDSSJ085829.88$+$022616.04 & 08:58:29.88 & $+$02:26:16.04 & $0.12\pm0.04$ & 18.3 & DIS & 1800 & 2009-03-26 \\
SDSSJ090519.01$+$084933.70 & 09:05:19.01 & $+$08:49:33.70 & $0.36\pm0.09$ & 20.3 & DIS & 1800 & 2009-03-19 \\
SDSSJ090519.61$+$084932.22 & 09:05:19.61 & $+$08:49:32.22 & $0.27\pm0.18$ & 20.9 & MagE & 1200 & 2009-02-22 \\
SDSSJ090519.72$+$084914.02 & 09:05:19.72 & $+$08:49:14.02 & ...           & 22.2 & MagE & $2\times 1800$ & 2009-02-22 \\
SDSSJ091845.10$+$060202.93 & 09:18:45.10 & $+$06:02:02.93 & $0.19\pm0.04$ & 18.6 & MagE & $600+3\times 300$ & 2009-03-29 \\
SDSSJ091845.70$+$060220.57 & 09:18:45.70 & $+$06:02:20.57 & $0.37\pm0.13$ & 19.9 & MagE & 600 & 2009-03-28 \\
SDSSJ093252.25$+$073731.59 & 09:32:52.25 & $+$07:37:31.59 & $0.35\pm0.11$ & 20.1 & MagE & 600 & 2009-03-29 \\
SDSSJ093537.25$+$112410.66 & 09:35:37.25 & $+$11:24:10.66 & $0.24\pm0.08$ & 20.4 & MagE & 600 & 2009-03-29 \\
SDSSJ100810.61$+$014446.17 & 10:08:10.61 & $+$01:44:46.17 & $0.17\pm0.03$ & 17.7 & DIS & 900 & 2009-03-26 \\
SDSSJ100906.91$+$023557.81 & 10:09:06.91 & $+$02:35:57.81 & $0.29\pm0.07$ & 19.2 & MagE & 300 & 2009-03-28 \\
SDSSJ102220.71$+$013143.50 & 10:22:20.71 & $+$01:31:43.50 & $0.12\pm0.01$ & 17.4 & SDSS & ... & ... \\
SDSSJ103605.26$+$015654.88 & 10:36:05.26 & $+$01:56:54.88 & $0.31\pm0.06$ & 18.8 & DIS & 1800 & 2009-03-19 \\
SDSSJ103836.38$+$095143.68 & 10:38:36.38 & $+$09:51:43.68 & $0.38\pm0.12$ & 20.3 & MagE & 600 & 2009-03-29 \\
SDSSJ112016.63$+$093317.94 & 11:20:16.63 & $+$09:33:17.94 & $0.34\pm0.11$ & 20.5 & MagE & 600 & 2009-03-28 \\
SDSSJ113756.76$+$085022.38 & 11:37:56.76 & $+$08:50:22.38 & $0.35\pm0.11$ & 20.7 & MagE & 600 & 2009-03-29 \\
SDSSJ114144.83$+$080554.09 & 11:41:44.83 & $+$08:05:54.09 & $0.17\pm0.09$ & 19.3 & DIS & 1800 & 2009-03-19 \\
SDSSJ114145.14$+$080605.27 & 11:41:45.14 & $+$08:06:05.27 & $0.34\pm0.07$ & 20.0 & DIS & 1800 & 2009-03-19 \\
SDSSJ114830.94$+$021807.91 & 11:48:30.94 & $+$02:18:07.91 & $0.33\pm0.04$ & 19.8 & DIS & 1200 & 2009-03-26 \\
SDSSJ114831.01$+$021803.00 & 11:48:31.01 & $+$02:18:03.00 & $0.29\pm0.05$ & 19.6 & DIS & 1200 & 2009-03-26 \\
SDSSJ120931.61$+$004546.23 & 12:09:31.61 & $+$00:45:46.23 & $0.30\pm0.09$ & 20.5 & MagE & $2\times 1200$ & 2009-02-22 \\
SDSSJ121347.09$+$000141.26 & 12:13:47.09 & $+$00:01:41.26 & $0.26\pm0.09$ & 20.2 & MagE & 1200 & 2009-06-20 \\
SDSSJ121347.14$+$000136.62 & 12:13:47.14 & $+$00:01:36.62 & $0.26\pm0.06$ & 19.1 & SDSS & ... & ... \\
SDSSJ122115.84$-$020259.37 & 12:21:15.84 & $-$02:02:59.37 & $0.25\pm0.16$ & 20.6 & MagE & 600 & 2008-03-29 \\
SDSSJ125737.93$+$144802.20 & 12:57:37.93 & $+$14:48:02.20 & $0.30\pm0.09$ & 20.3 & MagE & 600 & 2009-03-29 \\
SDSSJ130555.49$+$014928.62 & 13:05:55.49 & $+$01:49:28.62 & $0.32\pm0.06$ & 19.2 & DIS & 1800 & 2009-04-01 \\
SDSSJ130557.05$+$014922.34 & 13:05:57.05 & $+$01:49:22.34 & $0.17\pm0.01$ & 17.7 & DIS & 1800 & 2009-04-01 \\
SDSSJ132757.22$+$101136.02 & 13:27:57.22 & $+$10:11:36.02 & $0.37\pm0.13$ & 20.5 & MagE & 600 & 2009-03-28 \\
SDSSJ132829.30$+$080003.17 & 13:28:29.30 & $+$08:00:03.17 & $0.20\pm0.07$ & 19.3 & DIS & 1800 & 2009-04-01 \\
SDSSJ132830.62$+$080005.22 & 13:28:30.62 & $+$08:00:05.22 & $0.22\pm0.05$ & 19.0 & DIS & 1800 & 2009-04-01 \\
SDSSJ132831.15$+$075923.90 & 13:28:31.15 & $+$07:59:23.90 & $0.35\pm0.08$ & 20.4 & DIS & 1800 & 2009-04-01 \\
SDSSJ132831.54$+$075943.00 & 13:28:31.54 & $+$07:59:43.00 & $0.35\pm0.09$ & 19.5 & MagE & 300 & 2009-03-29 \\
SDSSJ132832.74$+$075952.56 & 13:28:32.74 & $+$07:59:52.56 & $0.19\pm0.09$ & 19.3 & DIS & 1800 & 2009-04-01 \\
SDSSJ133905.86$+$002225.36 & 13:39:05.86 & $+$00:22:25.36 & $0.14\pm0.01$ & 17.5 & SDSS & ... & ... \\
SDSSJ140618.34$+$130143.61 & 14:06:18.34 & $+$13:01:43.61 & $0.14\pm0.05$ & 18.0 & DIS & 1200 & 2009-04-01 \\
SDSSJ140619.94$+$130105.23 & 14:06:19.94 & $+$13:01:05.23 & $0.23\pm0.09$ & 19.7 & MagE & 300 & 2009-03-29 \\
SDSSJ142600.05$-$001818.12 & 14:26:00.05 & $-$00:18:18.12 & $0.14\pm0.01$ & 15.9 & SDSS & ... & ... \\
SDSSJ143216.97$+$095522.23 & 14:32:16.97 & $+$09:55:22.23 & $0.37\pm0.07$ & 20.5 & MagE & 600 & 2009-03-28 \\
SDSSJ150339.62$+$064235.04 & 15:03:39.62 & $+$06:42:35.04 & $0.22\pm0.10$ & 20.3 & DIS & 1800 & 2009-04-01 \\
SDSSJ150340.15$+$064308.11 & 15:03:40.15 & $+$06:43:08.11 & $0.30\pm0.11$ & 20.5 & MagE & $300+600$ & 2009-03-29 \\
SDSSJ151228.25$-$011216.09 & 15:12:28.25 & $-$01:12:16.09 & $0.16\pm0.03$ & 19.0 & MagE & 600 & 2009-02-23 \\
SDSSJ153112.77$+$091119.72 & 15:31:12.77 & $+$09:11:19.72 & $0.17\pm0.11$ & 20.8 & MagE & 600 & 2009-03-28 \\
SDSSJ153113.01$+$091127.02 & 15:31:13.01 & $+$09:11:27.02 & $0.22\pm0.11$ & 20.6 & MagE & 600 & 2009-03-28 \\
SDSSJ153715.67$+$023056.39 & 15:37:15.67 & $+$02:30:56.39 & $0.23\pm0.09$ & 19.5 & MagE & 900 & 2009-06-20 \\
SDSSJ155336.77$+$053438.23 & 15:53:36.77 & $+$05:34:38.23 & $0.24\pm0.06$ & 19.2 & DIS & 1800 & 2009-04-02 \\
SDSSJ155556.54$-$003615.58 & 15:55:56.54 & $-$00:36:15.58 & $0.24\pm0.15$ & 20.9 & MagE & 600 & 2009-03-29 \\
SDSSJ160749.54$-$002228.42 & 16:07:49.54 & $-$00:22:28.42 & $0.32\pm0.09$ & 19.6 & MagE & 900 & 2009-06-20 \\
SDSSJ160906.36$+$071330.66 & 16:09:06.36 & $+$07:13:30.66 & $0.18\pm0.05$ & 18.9 & MagE & $2\times 300$ & 2009-03-28 \\
SDSSJ204303.53$-$010139.05 & 20:43:03.53 & $-$01:01:39.05 & $0.19\pm0.04$ & 18.7 & DIS & $2\times 1800$ & 2008-10-24 \\
SDSSJ204304.34$-$010137.91 & 20:43:04.34 & $-$01:01:37.91 & $0.13\pm0.07$ & 19.5 & DIS & $2\times 1800$ & 2008-10-24 \\
SDSSJ204431.32$+$011304.97 & 20:44:31.32 & $+$01:13:04.97 & $0.16\pm0.04$ & 18.7 & MagE & $300+2\times 600$ & 2008-09-24 \\
SDSSJ204431.87$+$011308.81 & 20:44:31.87 & $+$01:13:08.81 & $0.19\pm0.11$ & 20.4 & DIS & 1800 & 2008-11-17 \\
SDSSJ210230.86$+$094121.06 & 21:02:30.86 & $+$09:41:21.06 & $0.29\pm0.09$ & 21.0 & MagE & $2\times 600$ & 2008-09-24 \\
SDSSJ212938.98$-$063758.80 & 21:29:38.98 & $-$06:37:58.80 & $0.22\pm0.13$ & 20.9 & MagE & 300 & 2008-09-23 \\
SDSSJ221126.42$+$124459.93 & 22:11:26.42 & $+$12:44:59.93 & $0.31\pm0.05$ & 20.4 & MagE & 600 & 2008-09-23 \\
SDSSJ221526.04$+$011353.78 & 22:15:26.04 & $+$01:13:53.78 & $0.25\pm0.11$ & 20.2 & MagE & $2\times 300$ & 2008-09-23 \\
SDSSJ221526.88$+$011347.20 & 22:15:26.88 & $+$01:13:47.20 & $0.14\pm0.19$ & 21.2 & DIS & $3\times 1800$ & 2008-11-23 \\
SDSSJ222849.01$-$005640.04 & 22:28:49.01 & $-$00:56:40.04 & $0.28\pm0.10$ & 19.7 & MagE & $2\times 400$ & 2008-09-25 \\
SDSSJ223246.44$+$134655.34 & 22:32:46.44 & $+$13:46:55.34 & $0.22\pm0.07$ & 19.1 & MagE & $2\times 300$ & 2008-09-23 \\
SDSSJ223316.34$+$133315.37 & 22:33:16.34 & $+$13:33:15.37 & $0.17\pm0.04$ & 18.8 & MagE & $200+300$ & 2008-09-25 \\
SDSSJ223359.74$-$003320.83 & 22:33:59.74 & $-$00:33:20.83 & $0.12\pm0.09$ & 19.8 & MagE & 1200 & 2009-06-21 \\
SDSSJ224704.01$-$081601.00 & 22:47:04.01 & $-$08:16:01.00 & $0.27\pm0.09$ & 19.8 & DIS & $3\times 1800$ & 2008-10-24 \\
SDSSJ230225.06$-$082156.65 & 23:02:25.06 & $-$08:21:56.65 & $0.32\pm0.07$ & 20.2 & MagE & $300+600$ & 2008-09-24 \\
SDSSJ230225.17$-$082159.07 & 23:02:25.17 & $-$08:21:59.07 & $0.49\pm0.29$ & 21.5 & MagE & $300+600$ & 2008-09-24 \\
SDSSJ230845.53$-$091445.97 & 23:08:45.53 & $-$09:14:45.97 & $0.26\pm0.09$ & 19.8 & MagE & $3\times 1200$ & 2008-09-25 \\
SDSSJ232812.79$-$090603.73 & 23:28:12.79 & $-$09:06:03.73 & $0.17\pm0.05$ & 18.6 & DIS & $2\times 1800$ & 2008-11-17 \\
SDSSJ234949.42$+$003542.34 & 23:49:49.42 & $+$00:35:42.34 & $0.23\pm0.09$ & 20.2 & MagE & $400+600$ & 2008-09-23 \\
\enddata
\end{deluxetable}
%\end{center}

%\begin{center}
%\begin{tiny}
\begin{deluxetable}{p{2in}ccccrr}
\tablewidth{0pc}
\tablecaption{Summary of the MagE Spectroscopic Observations of SDSS QSOs} 
\tabletypesize{\tiny}
\tablehead{ \colhead{ID} & \colhead{RA(J2000)} & \colhead{Dec(J2000)} & \colhead{$z_{\rm QSO}$} & \colhead{$u'$} & \colhead{Exptime} & \colhead{UT Date}}
\startdata
SDSSJ003340.21$-$005525.53 & 00:33:40.21 & $-$00:55:25.53 & 0.94 & 17.99 & $3\times 900$ & 2008-09-23 \nl
SDSSJ003407.34$-$085452.07 & 00:34:07.34 & $-$08:54:52.07 & 1.31 & 18.59 & $2\times 1200$ & 2008-09-24 \nl
SDSSJ003413.04$-$010026.86 & 00:34:13.04 & $-$01:00:26.86 & 1.29 & 17.33 & $2\times 600$ & 2008-09-23 \nl
SDSSJ010135.84$-$005009.08 & 01:01:35.84 & $-$00:50:09.08 & 1.01 & 19.31 & $2\times 1800$ & 2008-09-24 \nl
SDSSJ010156.32$-$084401.74 & 01:01:56.32 & $-$08:44:01.74 & 0.98 & 18.29 & $2\times 1800$ & 2008-09-25 \nl
SDSSJ010352.47$+$003739.79 & 01:03:52.47 & $+$00:37:39.79 & 0.70 & 18.36 & $3\times 1200$ & 2008-09-25 \nl
SDSSJ010508.14$-$005041.33 & 01:05:08.14 & $-$00:50:41.33 & 1.59 & 18.28 & $2\times 1200$ & 2008-09-25 \nl
SDSSJ021558.40$-$011135.79 & 02:15:58.40 & $-$01:11:35.79 & 0.84 & 17.85 & $2\times 1200$ & 2008-09-23 \nl
SDSSJ022950.32$-$074256.77 & 02:29:50.32 & $-$07:42:56.77 & 1.56 & 19.16 & $900+1200$ & 2008-09-25 \nl
SDSSJ024126.71$-$004526.25 & 02:41:26.71 & $-$00:45:26.25 & 0.72 & 18.36 & $2\times 900$ & 2008-09-24 \nl
SDSSJ032232.58$+$003649.13 & 03:22:32.58 & $+$00:36:49.13 & 1.59 & 19.50 & $2\times 1200$ & 2008-09-25 \nl
%SDSSJ033810.99$-$062041.25 & 03:38:10.99 & $-$06:20:41.25 & 0.55 & 18.99 & $2\times 1500$ & 2008-09-23 \nl
SDSSJ035242.12$+$001307.32 & 03:52:42.12 & $+$00:13:07.32 & 1.16 & 19.21 & $1400+1800$ & 2008-09-24 \nl
SDSSJ040404.08$-$060714.03 & 04:04:04.08 & $-$06:07:14.03 & 1.29 & 19.00 & $2\times 1800$ & 2009-02-22 \nl
SDSSJ074528.15$+$191952.68 & 07:45:28.15 & $+$19:19:52.68 & 0.69 & 17.95 & $2\times 1800$ & 2009-02-22 \nl
SDSSJ075001.85$+$161305.05 & 07:50:01.85 & $+$16:13:05.05 & 1.10 & 18.41 & $2\times 1800$ & 2009-03-28 \nl
SDSSJ075450.04$+$184952.79 & 07:54:50.04 & $+$18:49:52.79 & 0.81 & 18.37 & $2\times 1800$ & 2009-02-23 \nl
SDSSJ075525.51$+$172836.59 & 07:55:25.51 & $+$17:28:36.59 & 1.29 & 18.38 & $2\times 1800$ & 2009-02-22 \nl
SDSSJ080004.56$+$184935.15 & 08:00:04.56 & $+$18:49:35.15 & 1.29 & 17.99 & $1500+1800$ & 2009-03-29 \nl
SDSSJ082340.18$+$074801.68 & 08:23:40.18 & $+$07:48:01.68 & 0.84 & 18.33 & $2\times 1800$ & 2009-03-28 \nl
SDSSJ083220.74$+$043416.78 & 08:32:20.74 & $+$04:34:16.78 & 1.51 & 18.25 & $3\times 1800$ & 2008-02-03 \nl
SDSSJ084119.78$+$012621.75 & 08:41:19.78 & $+$01:26:21.75 & 1.48 & 18.20 & $2\times 1800$ & 2008-02-02 \nl
SDSSJ084456.06$+$004708.95 & 08:44:56.06 & $+$00:47:08.95 & 1.31 & 18.45 & $2\times 1800$ & 2009-02-23 \nl
SDSSJ085826.93$+$022604.49 & 08:58:26.93 & $+$02:26:04.49 & 1.51 & 18.19 & $1200+1800$ & 2008-02-02 \nl
SDSSJ090519.70$+$084917.32 & 09:05:19.70 & $+$08:49:17.32 & 1.43 & 18.45 & $2\times 1800$ & 2009-02-22 \nl
SDSSJ091845.91$+$060226.09 & 09:18:45.91 & $+$06:02:26.09 & 0.79 & 18.31 & $2\times 1800$ & 2009-03-28 \nl
SDSSJ093251.82$+$073729.11 & 09:32:51.82 & $+$07:37:29.11 & 0.90 & 17.87 & $2\times 1800$ & 2009-03-29 \nl
SDSSJ093536.98$+$112408.03 & 09:35:36.98 & $+$11:24:08.03 & 0.85 & 18.30 & $600+1800$ & 2009-03-29 \nl
SDSSJ100807.51$+$014448.97 & 10:08:07.51 & $+$01:44:48.97 & 1.33 & 18.25 & $2\times 1800$ & 2009-02-23 \nl
SDSSJ100906.36$+$023555.31 & 10:09:06.36 & $+$02:35:55.31 & 1.10 & 17.20 & $2\times 1200$  & 2009-03-28\nl
SDSSJ102218.98$+$013218.82 & 10:22:18.98 & $+$01:32:18.82 & 0.79 & 16.92 & $900+1800$ & 2009-02-22 \nl
SDSSJ103607.51$+$015659.14 & 10:36:07.51 & $+$01:56:59.14 & 1.86 & 19.07 & $2\times 1800$ & 2008-01-31 \nl
SDSSJ103836.50$+$095138.85 & 10:38:36.50 & $+$09:51:38.85 & 1.02 & 18.36 & $2\times 1200$ & 2009-03-29 \nl
SDSSJ112016.66$+$093323.53 & 11:20:16.66 & $+$09:33:23.53 & 1.10 & 18.02 & $2\times 1800$ & 2009-03-28 \nl
SDSSJ113757.02$+$085017.21 & 11:37:57.02 & $+$08:50:17.21 & 1.10 & 18.38 & $1200$ & 2009-03-29 \nl
SDSSJ114144.62$+$080614.79 & 11:41:44.62 & $+$08:06:14.79 & 1.08 & 18.37 & $1500+1800$ & 2009-02-23 \nl
SDSSJ114830.12$+$021829.78 & 11:48:30.12 & $+$02:18:29.78 & 1.22 & 18.00 & $2\times 1700$ & 2008-01-31 \nl
SDSSJ120932.26$+$004555.92 & 12:09:32.26 & $+$00:45:55.92 & 1.44 & 18.42 & $2\times 1800$ & 2009-02-22 \nl
SDSSJ121347.52$+$000129.99 & 12:13:47.52 & $+$00:01:29.99 & 0.96 & 18.35 & $2\times 1800$ & 2009-06-21 \nl
SDSSJ122115.34$-$020253.39 & 12:21:15.34 & $-$02:02:53.39 & 0.79 & 17.82 & $1200$ & 2009-03-29 \nl
SDSSJ125739.22$+$144806.26 & 12:57:39.22 & $+$14:48:06.26 & 0.82 & 18.34 & $1200$ & 2009-03-29 \nl
SDSSJ130554.17$+$014929.82 & 13:05:54.17 & $+$01:49:29.82 & 0.73 & 18.00 & $1200+1800$ & 2009-02-23 \nl
SDSSJ132757.41$+$101141.78 & 13:27:57.41 & $+$10:11:41.78 & 1.37 & 18.14 & $2\times 1200$ & 2009-03-28 \nl
SDSSJ132831.08$+$075942.01 & 13:28:31.08 & $+$07:59:42.01 & 1.33 & 18.26 & $1200$ & 2009-03-29 \nl
SDSSJ133904.34$+$002221.92 & 13:39:04.34 & $+$00:22:21.92 & 1.15 & 17.98 & $1200+1800$ & 2009-06-21 \nl
SDSSJ140619.61$+$130106.82 & 14:06:19.61 & $+$13:01:06.82 & 1.02 & 18.35 & $1200$ & 2009-03-29 \nl
SDSSJ142556.40$-$001818.79 & 14:25:56.40 & $-$00:18:18.79 & 1.15 & 18.96 & $1500+3000$ & 2008-07-24 \nl
SDSSJ143216.78$+$095519.29 & 14:32:16.78 & $+$09:55:19.29 & 0.77 & 17.67 & $1200$ & 2009-03-28 \nl
SDSSJ150339.98$+$064259.96 & 15:03:39.98 & $+$06:42:59.96 & 0.94 & 18.26 & $1200$ & 2009-03-29 \nl
SDSSJ151228.82$-$011223.12 & 15:12:28.82 & $-$01:12:23.12 & 1.17 & 18.39 & $1200+1800$ & 2009-02-23 \nl
SDSSJ153112.98$+$091138.78 & 15:31:12.98 & $+$09:11:38.78 & 1.23 & 18.19 & $600+1200$ & 2009-03-28 \nl
SDSSJ153715.34$+$023049.73 & 15:37:15.34 & $+$02:30:49.73 & 0.48 & 17.65 & $2\times 2400$ & 2009-06-20 \nl
SDSSJ155336.46$+$053423.97 & 15:53:36.46 & $+$05:34:23.97 & 1.14 & 18.38 & $2\times 1800$ & 2009-06-21 \nl
SDSSJ155557.07$-$003608.41 & 15:55:57.07 & $-$00:36:08.41 & 0.76 & 18.22 & $1200$ & 2009-03-29 \nl
SDSSJ160749.34$-$002219.86 & 16:07:49.34 & $-$00:22:19.86 & 1.30 & 18.05 & $2\times 1800$ & 2009-06-20 \nl
SDSSJ160905.42$+$071337.29 & 16:09:05.42 & $+$07:13:37.29 & 0.70 & 18.10 & $2\times 1200$ & 2009-03-28 \nl
%SDSSJ203952.18$-$002447.12 & 20:39:52.18 & $-$00:24:47.12 & 0.61 & 18.39 & $3\times 2400$ & 2009-06-21 \nl
SDSSJ204303.55$-$010126.05 & 20:43:03.55 & $-$01:01:26.05 & 1.19 & 17.62 & $2\times 1200$ & 2008-07-27\nl
SDSSJ204431.46$+$011312.43 & 20:44:31.46 & $+$01:13:12.43 & 0.98 & 19.07 & $3\times 1800$ & 2008-09-24 \nl
SDSSJ210230.72$+$094125.08 & 21:02:30.72 & $+$09:41:25.08 & 0.79 & 19.19 & $2\times 1200$ & 2008-09-24 \nl
%SDSSJ212034.87$-$003103.85 & 21:20:34.87 & $-$00:31:03.85 & 1.48 & 17.91 & $2\times 1800$ & 2009-06-21 \nl
SDSSJ212938.59$-$063801.85 & 21:29:38.59 & $-$06:38:01.85 & 1.01 & 18.08 & $900+1200$ & 2008-09-23 \nl
SDSSJ221126.76$+$124458.16 & 22:11:26.76 & $+$12:44:58.16 & 0.49 & 18.16 & $2\times 900$ & 2008-09-23 \nl
SDSSJ221526.74$+$011356.47 & 22:15:26.74 & $+$01:13:56.47 & 1.26 & 19.21 & $2\times 1200$ & 2008-09-23 \nl
SDSSJ222849.20$-$005630.89 & 22:28:49.20 & $-$00:56:30.89 & 1.28 & 18.96 & $900+1800$ & 2008-09-25 \nl
SDSSJ223246.80$+$134702.04 & 22:32:46.80 & $+$13:47:02.04 & 1.55 & 18.45 & $2\times 1200$ & 2008-09-23 \nl
SDSSJ223316.87$+$133309.90 & 22:33:16.87 & $+$13:33:09.90 & 1.54 & 18.59 & $2\times 1800$ & 2008-09-25 \nl
SDSSJ223359.93$-$003315.79 & 22:33:59.93 & $-$00:33:15.79 & 1.21 & 17.42 & $2\times 1800$ & 2009-06-21 \nl
SDSSJ224704.78$-$081617.54 & 22:47:04.78 & $-$08:16:17.54 & 0.97 & 17.54 & $1200+3000$ & 2008-07-25 \nl
SDSSJ230225.49$-$082154.12 & 23:02:25.49 & $-$08:21:54.12 & 1.01 & 19.13 & $2\times 1200$ & 2008-09-24 \nl
SDSSJ230845.60$-$091449.45 & 23:08:45.60 & $-$09:14:49.45 & 0.89 & 19.27 & $3\times 1200$ & 2008-09-25 \nl
SDSSJ232812.91$-$090522.56 & 23:28:12.91 & $-$09:05:22.56 & 1.11 & 17.71 & $600+2400$  & 2008-07-25 \nl
SDSSJ234949.61$+$003535.39 & 23:49:49.61 & $+$00:35:35.39 & 1.24 & 18.26 & $1200+1500$ & 2008-09-23 \nl
%SDSSJ235321.82$-$011528.18 & 23:53:21.82 & $-$01:15:28.18 & 0.99 & 16.92 & $900$ & 2009-06-21 \nl
\enddata
\end{deluxetable}
%\end{tiny}
%\end{center}

%\begin{center}
%\begin{tiny}
\begin{deluxetable}{lrrccccrccl}
%\multicolumn{19}{c}{Table 1.4} \\
\tablewidth{0pc}
\tablecaption{Galaxies and Absorption Systems}
\tabletypesize{\tiny}
\tablehead{\multicolumn{8}{c}{Galaxies} & & \multicolumn{2}{c}{Absorption Systems} \\
\cline{1-8}
\cline{10-11}
& \multicolumn{1}{c}{$\Delta \alpha$} & \multicolumn{1}{c}{$\Delta \delta$} & &
\multicolumn{1}{c}{$\rho$} & & \colhead{$M_B$} & & & & \colhead{$W(2796)$\tablenotemark{a}} \\
\multicolumn{1}{c}{ID} & \multicolumn{1}{c}{(arcsec)} &
\multicolumn{1}{c}{(arcsec)} & \colhead{$z_{\rm gal}$} & \colhead{($h^{-1}$ kpc)} & \colhead{$r'$} &
\colhead{$-5 \log h$} & \colhead{$B_{AB}-R_{AB}$} & & \colhead{$z_{\rm abs}$} & \colhead{(\AA)} \\
\multicolumn{1}{c}{(1)} & \multicolumn{1}{c}{(2)} & \multicolumn{1}{c}{(3)} &
\colhead{(4)} & \colhead{(5)} & \colhead{(6)} & \colhead{(7)} & \colhead{(8)} &  & \colhead{(9)} & \colhead{(10)} }
\startdata
SDSSJ003339.85$-$005522.36 &  $-5.4$ &   $3.2$ & 0.2124 &  15.2 & 18.82 & $-20.11$ & $0.25$ & & 0.2121 & $1.05\pm0.03$ \nl
SDSSJ003407.78$-$085453.28 &   $6.5$ &  $-1.2$ & 0.3617 &  23.5 & 21.37 & $-18.89$ & $0.79$ & & 0.3616 & $0.48\pm0.05$ \nl
SDSSJ003412.85$-$010019.79 &  $-2.8$ &   $7.1$ & 0.2564 &  21.3 & 20.16 & $-18.94$ & $1.11$ & & 0.2564 & $0.61\pm0.06$ \nl
SDSSJ003414.49$-$005927.49 &  $21.7$ &  $59.4$ & 0.1202 &  96.0 & 17.24 & $-19.80$ & $1.14$ & & 0.1202 & $<0.22$ \nl
SDSSJ010136.52$-$005016.44 &  $10.2$ &  $-7.4$ & 0.2615 &  35.9 & 19.43 & $-19.71$ & $1.15$ & & 0.2615 & $<0.07$ \nl
SDSSJ010155.80$-$084408.74 &  $-7.7$ &  $-7.0$ & 0.1588 &  20.0 & 20.19 & $-17.70$ & $0.65$ & & 0.1586 & $0.36\pm0.03$ \nl
SDSSJ010351.82$+$003740.77 &  $-9.7$ &   $1.0$ & 0.3515 &  34.0 & 21.38 & $-18.85$ & $0.88$ & & 0.3508 & $0.38\pm0.03$ \nl
SDSSJ021558.84$-$011131.23 &   $6.6$ &   $4.6$ & 0.2103 &  19.3 & 19.35 & $-19.35$ & $0.84$ & & 0.2108 & $0.77\pm0.05$ \nl
SDSSJ022949.97$-$074255.88 &  $-5.2$ &   $0.9$ & 0.3866 &  19.5 & 20.94 & $-19.47$ & $0.62$ & & 0.3861 & $1.74\pm0.04$ \nl
SDSSJ024127.75$-$004517.04 &  $15.6$ &   $9.2$ & 0.1765 &  37.9 & 18.69 & $-19.39$ & $1.26$ & & ... & ... \nl
SDSSJ032230.27$+$003712.72 & $-34.6$ &  $23.6$ & 0.1833 &  90.4 & 17.55 & $-21.02$ & $0.55$ & & ... & ... \nl
SDSSJ032232.55$+$003644.68 &  $-0.4$ &  $-4.5$ & 0.2185 &  11.1 & 21.27 & $-17.48$ & $0.85$ & & 0.2183 & $1.31\pm0.12$ \nl
SDSSJ035241.99$+$001317.13 &  $-1.9$ &   $9.8$ & 0.3671 &  35.9 & 20.37 & $-19.87$ & $0.72$ & & 0.3677 & $1.45\pm0.05$ \nl
SDSSJ040404.51$-$060709.46 &   $6.4$ &   $4.6$ & 0.2387 &  20.8 & 20.20 & $-18.89$ & $0.62$ & & ... & ... \nl
SDSSJ075001.34$+$161301.92 &  $-7.3$ &  $-3.1$ & 0.1466 &  14.3 & 20.65 & $-17.19$ & $0.66$ & & 0.1469 & $0.26\pm0.08$ \nl
SDSSJ075450.11$+$185005.28 &   $1.0$ &  $12.5$ & 0.2856 &  37.8 & 19.77 & $-19.69$ & $1.08$ & & 0.2856 & $<0.04$ \nl
SDSSJ075525.13$+$172825.79 &  $-5.4$ & $-10.8$ & 0.2541 &  33.5 & 19.34 & $-19.82$ & $1.02$ & & 0.2546 & $0.51\pm0.02$ \nl
SDSSJ080005.11$+$184933.31 &   $7.8$ &  $-1.8$ & 0.2544 &  22.2 & 19.94 & $-19.33$ & $0.48$ & & 0.2536 & $0.30\pm0.04$ \nl
SDSSJ082340.56$+$074751.07 &   $5.6$ & $-10.6$ & 0.1864 &  26.3 & 18.54 & $-19.67$ & $1.00$ & & 0.1863 & $0.37\pm0.04$ \nl
SDSSJ084120.59$+$012628.85 &  $12.1$ &   $7.1$ & 0.4091 &  53.7 & 20.05 & $-20.54$ & $0.81$ & & 0.4084 & $0.10\pm0.02$ \nl
SDSSJ084455.58$+$004718.15 &  $-7.2$ &   $9.2$ & 0.1551 &  22.0 & 18.86 & $-19.10$ & $0.72$ & & 0.1554 & $0.40\pm0.05$ \nl
SDSSJ085829.88$+$022616.04 &  $44.2$ &  $11.6$ & 0.1097 &  64.0 & 18.34 & $-18.81$ & $0.75$ & & 0.1097 & $<0.06$ \nl
SDSSJ090519.01$+$084933.70 & $-10.2$ &  $16.4$ & 0.3856 &  71.1 & 20.33 & $-20.07$ & $-0.13$& & 0.3856 & $<0.04$ \nl
SDSSJ090519.61$+$084932.22 &  $-1.3$ &  $14.9$ & 0.4545 &  60.7 & 20.96 & $-19.96$ & $0.90$ & & 0.4545 & $<0.04$ \nl
SDSSJ090519.72$+$084914.02 &   $0.3$ &  $-3.3$ & 0.1499 &   6.1 & 22.30 & $-16.36$ & $0.66$ & & 0.1501 & $0.82\pm0.10$ \nl
SDSSJ091845.10$+$060202.93 & $-12.1$ & $-23.2$ & 0.1849 &  56.7 & 18.70 & $-19.43$ & $1.19$ & & 0.1849 & $<0.07$ \nl
SDSSJ093252.25$+$073731.59 &   $6.4$ &   $2.5$ & 0.3876 &  25.3 & 20.20 & $-20.23$ & $0.29$ & & 0.3876 & $1.10\pm0.02$ \nl
SDSSJ093537.25$+$112410.66 &   $4.0$ &   $2.6$ & 0.2808 &  14.2 & 20.41 & $-19.10$ & $0.34$ & & 0.2811 & $0.79\pm0.04$ \nl
SDSSJ100810.61$+$014446.17 &  $46.5$ &  $-2.8$ & 0.2173 & 114.8 & 17.73 & $-21.07$ & $0.80$ & & 0.2173 & $<0.20$ \nl
SDSSJ100906.91$+$023557.81 &   $8.2$ &   $2.5$ & 0.2523 &  23.7 & 19.21 & $-19.85$ & $1.13$ & & 0.2521 & $0.10\pm0.01$ \nl
SDSSJ102220.71$+$013143.50 &  $25.9$ & $-35.3$ & 0.1369 &  74.3 & 17.45 & $-19.99$ & $0.99$ & & 0.1369 & $<0.11$ \nl
SDSSJ103605.26$+$015654.88 & $-33.7$ &  $-4.3$ & 0.3571 & 119.2 & 18.89 & $-21.29$ & $0.59$ & & 0.3571 & $<0.02$ \nl
SDSSJ103836.38$+$095143.68 &  $-1.8$ &   $4.8$ & 0.1742 &  10.7 & 20.39 & $-18.03$ & $0.40$ & & 0.1744 & $1.04\pm0.06$ \nl
SDSSJ112016.63$+$093317.94 &  $-0.4$ &  $-5.6$ & 0.4933 &  23.8 & 20.59 & $-20.63$ & $0.90$ & & 0.4933 & $2.14\pm0.03$ \nl
SDSSJ113756.76$+$085022.38 &  $-3.9$ &   $5.2$ & 0.3356 &  21.7 & 20.75 & $-19.25$ & $1.11$ & & 0.3360 & $0.91\pm0.06$ \nl
SDSSJ114144.83$+$080554.09 &   $3.1$ & $-20.7$ & 0.2290 &  53.7 & 19.37 & $-19.60$ & $0.80$ & & 0.2286 & $0.31\pm0.03$ \nl
SDSSJ114145.14$+$080605.27 &   $7.7$ &  $-9.5$ & 0.3583 &  43.1 & 20.09 & $-20.08$ & $1.14$ & & 0.3585 & $0.49\pm0.02$ \nl
SDSSJ120931.61$+$004546.23 &  $-9.7$ &  $-9.7$ & 0.2533 &  38.0 & 20.50 & $-18.76$ & $0.94$ & & 0.2533 & $<0.06$ \nl
SDSSJ122115.84$-$020259.37 &   $7.5$ &  $-6.0$ & 0.0934 &  11.7 & 20.65 & $-16.22$ & $0.51$ & & ... & ... \nl
SDSSJ125737.93$+$144802.20 &  $-4.2$ &  $-4.1$ & 0.4648 &  24.0 & 20.32 & $-20.67$ & $1.06$ & & 0.4644 & $0.12\pm0.02$ \nl
SDSSJ130555.49$+$014928.62 &  $19.8$ &  $-1.2$ & 0.2258 &  50.3 & 19.29 & $-19.58$ & $0.94$ & & 0.2258 & $<0.04$ \nl
SDSSJ130557.05$+$014922.34 &  $43.2$ &  $-7.5$ & 0.1747 &  90.9 & 17.70 & $-20.30$ & $1.16$ & & 0.1740 & $0.45\pm0.03$ \nl
SDSSJ132757.22$+$101136.02 &  $-2.8$ &  $-5.8$ & 0.2557 &  17.8 & 20.60 & $-18.61$ & $0.45$ & & 0.2553 & $0.65\pm0.04$ \nl
SDSSJ132831.54$+$075943.00 &   $6.8$ &   $1.0$ & 0.3323 &  23.1 & 19.50 & $-20.46$ & $1.07$ & & 0.3326 & $0.59\pm0.04$ \nl
SDSSJ132832.74$+$075952.56 &  $24.7$ &  $10.6$ & 0.2358 &  70.3 & 19.34 & $-19.75$ & $0.57$ & & 0.2362 & $0.21\pm0.05$ \nl
SDSSJ133905.86$+$002225.36 &  $22.8$ &   $3.4$ & 0.1438 &  40.8 & 17.55 & $-19.92$ & $1.17$ & & 0.1438 & $<0.22$ \nl
SDSSJ140618.34$+$130143.61 & $-18.6$ &  $36.8$ & 0.1748 &  85.6 & 18.08 & $-20.04$ & $1.00$ & & 0.1748 & $<0.11$ \nl
SDSSJ140619.94$+$130105.23 &   $4.8$ &  $-1.6$ & 0.2220 &  12.7 & 19.76 & $-19.06$ & $0.51$ & & 0.2222 & $0.96\pm0.06$ \nl
SDSSJ142600.05$-$001818.12 &  $54.7$ &   $0.7$ & 0.1382 &  93.6 & 15.96 & $-21.43$ & $1.13$ & & 0.1382 & $<0.19$ \nl
SDSSJ143216.97$+$095522.23 &   $2.8$ &   $2.9$ & 0.3293 &  13.5 & 20.59 & $-19.31$ & $0.97$ & & 0.3296 & $2.36\pm0.04$ \nl
SDSSJ150339.62$+$064235.04 &  $-5.4$ & $-24.9$ & 0.2333 &  66.3 & 20.35 & $-18.73$ & $0.50$ & & 0.2333 & $<0.06$ \nl
SDSSJ150340.15$+$064308.11 &   $2.5$ &   $8.2$ & 0.1809 &  18.2 & 20.57 & $-17.47$ & $0.97$ & & 0.1809 & $<0.11$ \nl
SDSSJ151228.25$-$011216.09 &  $-8.5$ &   $7.0$ & 0.1284 &  17.8 & 19.08 & $-18.51$ & $0.59$ & & 0.1284 & $0.94\pm0.16$ \nl
SDSSJ153112.77$+$091119.72 &  $-3.1$ & $-19.1$ & 0.3265 &  63.8 & 20.89 & $-19.03$ & $0.70$ & & 0.3265 & $<0.04$ \nl
SDSSJ153113.01$+$091127.02 &   $0.4$ & $-11.8$ & 0.2659 &  33.7 & 20.67 & $-18.60$ & $1.08$ & & 0.2660 & $0.31\pm0.03$ \nl
SDSSJ153715.67$+$023056.39 &   $4.9$ &   $6.7$ & 0.2151 &  20.3 & 19.51 & $-19.41$ & $0.56$ & & 0.2151 & $0.80\pm0.02$ \nl
SDSSJ155336.77$+$053438.23 &   $4.6$ &  $14.3$ & 0.3227 &  49.0 & 19.22 & $-20.66$ & $0.56$ & & 0.3240 & $0.71\pm0.01$ \nl
SDSSJ155556.54$-$003615.58 &  $-7.9$ &  $-7.2$ & 0.3006 &  33.5 & 20.91 & $-18.85$ & $1.01$ & & 0.3006 & $<0.04$ \nl
SDSSJ160749.54$-$002228.42 &   $3.0$ &  $-8.6$ & 0.3985 &  34.1 & 19.63 & $-20.88$ & $0.84$ & & 0.3993 & $0.80\pm0.01$ \nl
SDSSJ160906.36$+$071330.66 &  $14.0$ &  $-6.6$ & 0.2075 &  36.8 & 18.93 & $-19.63$ & $1.10$ & & 0.2065 & $<0.08$ \nl
SDSSJ204303.53$-$010139.05 &  $-0.3$ & $-13.0$ & 0.2356 &  34.1 & 18.73 & $-20.38$ & $0.55$ & & 0.2350 & $1.24\pm0.05$ \nl
SDSSJ204304.34$-$010137.91 &  $11.8$ & $-11.9$ & 0.1329 &  39.9 & 19.56 & $-18.03$ & $0.60$ & & 0.1329 & $<0.19$ \nl
SDSSJ210230.86$+$094121.06 &   $2.1$ &  $-4.0$ & 0.3565 &  15.8 & 21.05 & $-19.12$ & $1.53$ & & 0.3563 & $0.71\pm0.04$ \nl
SDSSJ212938.98$-$063758.80 &   $5.8$ &   $3.0$ & 0.2782 &  19.4 & 20.94 & $-18.49$ & $0.92$ & & 0.2779 & $0.58\pm0.03$ \nl
SDSSJ221126.42$+$124459.93 &  $-5.0$ &   $1.8$ & 0.4872 &  22.2 & 20.46 & $-20.51$ & $0.55$ & & 0.4840 & $0.40\pm0.02$ \nl
SDSSJ221526.04$+$011353.78 & $-10.5$ &  $-2.7$ & 0.3203 &  35.4 & 20.26 & $-19.61$ & $1.05$ & & 0.3201 & $0.40\pm0.05$ \nl
SDSSJ221526.88$+$011347.20 &   $2.1$ &  $-9.3$ & 0.1952 &  21.5 & 21.25 & $-17.48$ & $0.51$ & & 0.1952 & $<0.15$ \nl
SDSSJ222849.01$-$005640.04 &  $-2.8$ &  $-9.2$ & 0.2410 &  25.5 & 19.80 & $-19.27$ & $0.88$ & & ... & ... \nl
SDSSJ223246.44$+$134655.34 &  $-5.2$ &  $-6.7$ & 0.3221 &  27.9 & 19.19 & $-20.68$ & $1.00$ & & 0.3225 & $0.92\pm0.05$ \nl
SDSSJ223316.34$+$133315.37 &  $-7.7$ &   $5.5$ & 0.2138 &  23.1 & 18.85 & $-20.08$ & $0.69$ & & 0.2139 & $1.36\pm0.06$ \nl
SDSSJ223359.74$-$003320.83 &  $-2.8$ &  $-5.0$ & 0.1162 &   8.5 & 19.86 & $-17.37$ & $0.78$ & & 0.1162 & $1.11\pm0.09$ \nl
SDSSJ224704.01$-$081601.00 & $-11.4$ &  $16.5$ & 0.4270 &  78.7 & 19.84 & $-20.88$ & $1.34$ & & 0.4270 & $<0.04$ \nl
SDSSJ230225.06$-$082156.65 &  $-6.4$ &  $-2.5$ & 0.3618 &  24.2 & 20.27 & $-19.92$ & $0.97$ & & 0.3620 & $2.02\pm0.06$ \nl
SDSSJ230225.17$-$082159.07 &  $-4.7$ &  $-4.9$ & 0.2146 &  16.8 & 21.59 & $-17.49$ & $0.53$ & & ... & ... \nl
SDSSJ230845.53$-$091445.97 &  $-1.0$ &   $3.5$ & 0.2147 &   8.9 & 19.89 & $-18.85$ & $0.96$ & & 0.2139 & $0.43\pm0.07$ \nl
SDSSJ232812.79$-$090603.73 &  $-1.8$ & $-41.2$ & 0.1148 &  60.1 & 18.64 & $-18.73$ & $0.50$ & & 0.1148 & $<0.24$ \nl
SDSSJ234949.42$+$003542.34 &  $-2.8$ &   $7.0$ & 0.2778 &  22.2 & 20.29 & $-19.14$ & $0.96$ & & 0.2776 & $0.35\pm0.02$ \nl
\hline 
\multicolumn{10}{c}{'Group' Galaxies} \\
\hline
SDSSJ003339.66$-$005518.36 &  $-8.2$ &   $7.2$ & 0.1760 &  22.8 & 19.66 & $-18.36$ & $1.12$ & & 0.1759 & $0.19\pm0.04$ \nl
SDSSJ003341.47$-$005522.79 &  $18.9$ &   $2.7$ & 0.1758 &  39.8 & 20.60 & $-17.91$ & $0.98$ & & ... & ... \nl
SDSSJ074527.07$+$191959.90 & $-15.0$ &  $ 7.5$ & 0.4582 &  68.9 & 19.71 & $-21.22$ & $1.12$ & & 0.4549 & $0.65\pm0.10$ \nl
SDSSJ074527.22$+$192003.88 & $-13.2$ &  $11.2$ & 0.4582 &  70.4 & 20.21 & $-20.72$ & $0.82$ & & ... & ... \nl
SDSSJ083218.55$+$043337.81 & $-32.7$ & $-39.0$ & 0.1681 & 102.4 & 18.14 & $-19.86$ & $1.03$ & & 0.1684 & $0.20\pm0.04$ \nl
SDSSJ083218.77$+$043346.58 & $-29.5$ & $-30.2$ & 0.1678 &  84.7 & 17.47 & $-20.49$ & $0.98$ & & ... & ... \nl
SDSSJ083221.60$+$043359.74 &  $12.9$ & $-17.0$ & 0.1693 &  43.2 & 19.41 & $-18.91$ & $0.50$ & & ... & ... \nl
SDSSJ091845.70$+$060220.57 &  $-3.1$ &  $-5.5$ & 0.7967 &  33.4 & 19.94 & $-22.55$ &$-1.36$ & & ... & $<0.02$ \nl
SDSSJ114830.94$+$021807.91 &  $12.3$ & $-21.9$ & 0.3206 &  81.9 & 19.82 & $-19.97$ & $1.06$ & & 0.3215 & $0.53\pm0.02$ \nl
SDSSJ114831.01$+$021803.00 &  $13.4$ & $-26.9$ & 0.3206 &  98.0 & 19.62 & $-20.17$ & $1.06$ & & ... & ... \nl
SDSSJ121347.09$+$000141.26 &  $-6.4$ &  $11.3$ & 0.2258 &  33.0 & 20.26 & $-18.67$ & $1.15$ & & 0.2258 & $0.54\pm0.08$ \nl
SDSSJ121347.14$+$000136.62 &  $-5.7$ &   $6.6$ & 0.2259 &  22.2 & 19.15 & $-19.57$ & $1.10$ & & ... & ... \nl
SDSSJ132829.30$+$080003.17 & $-26.4$ &  $21.2$ & 0.2549 &  94.1 & 19.32 & $-19.92$ & $0.80$ & & 0.2545 & $0.79\pm0.03$ \nl
SDSSJ132830.62$+$080005.22 &  $-6.8$ &  $23.2$ & 0.2537 &  67.0 & 19.03 & $-20.08$ & $0.95$ & & ... & ... \nl
SDSSJ132831.15$+$075923.90 &   $1.0$ & $-18.1$ & 0.2537 &  50.2 & 20.41 & $-18.69$ & $1.10$ & & ... & ... \nl
SDSSJ204431.32$+$011304.97 &  $-2.1$ &  $-7.5$ & 0.1927 &  17.4 & 18.76 & $-19.47$ & $1.21$ & & 0.1927 & $0.50\pm0.08$ \nl
SDSSJ204431.87$+$011308.81 &   $6.1$ &  $-3.6$ & 0.1921 &  16.0 & 20.44 & $-18.00$ & $0.77$ & & ... & ... \nl
\enddata
\tablenotetext{a}{Constraints for the strength of the corresponding Mg\,II absorption at the redshift of each galaxy.  No detections are expressed as 2-$\sigma$ upper limits to the underlying absorption strength of cool gas.  Galaxies that do not have constraints for \ewr\ due to contamination or are members of a 'group' are indicated with '...'.}
\end{deluxetable}
%s\end{tiny}
%\end{center}

%\clearpage

%\begin{figure}
%\begin{center}
%\includegraphics[scale=0.6]{w_rhomz_m1.pdf}
%\includegraphics[scale=0.75]{f3.eps}
%\caption{Comparison of $W_r(2796)$ versus impact parameter accouting
%  for scaling by $B$-band luminosity and redshift.  The reference
%  redshift $z_0$ is adopted to be the median redshift of the galaxy
%  sample $\langle z\rangle=0.25$.  Symbols are the same as in Figure
%  8. The solid line is the best-fit power-law model, excluding four
%  outliers marked in dotted circles.  The errorbar in the lower-left
%  corner indicates the intrinsic scatter estimated based on the
%  likelihood analysis.}
%\end{center} 
%\end{figure}

\end{document}